%% file: main.tex
\pdfoutput=1
\documentclass{article} %
\usepackage{iclr2027_conference,times}

\input{math_commands.tex}

\usepackage{hyperref}
\usepackage{url}

\usepackage{booktabs}
\usepackage{amsfonts}
\usepackage{nicefrac}
\usepackage{microtype}
\usepackage{xcolor}
\usepackage{graphicx}
\usepackage{amsmath}
\usepackage{amssymb}
\usepackage{algorithm}
\usepackage{algpseudocode}
\usepackage{subcaption}
\usepackage{multirow}
\usepackage{comment}
\usepackage{tikz}
\usepackage{xcolor}
\usetikzlibrary{positioning, arrows.meta, fit, backgrounds, calc, shapes.geometric}

\iclrfinalcopy

\title{Bio-Inspired Auditory Processing: Coupling Parallel Resonate-and-Fire Dynamics with Ternary Recurrent Networks for Speech Recognition}

\author{
  \centerline{
    \textbf{Pavlos Poulos}$^{1,2}$ \quad 
    \textbf{Katelyn Sadorf}$^{3,4}$ \quad 
    \textbf{Themos Stafylakis}$^{1,2}$ \quad 
    \textbf{Costas A. Anastassiou}$^{2,3,4}$
  } \\[1.5ex]
  \centerline{\normalfont $^1$Department of Informatics, Athens University of Economics and Business, Greece} \\
  \centerline{\normalfont $^2$Archimedes AI, Athena Research Center, Greece} \\
  \centerline{\normalfont $^3$California Institute of Technology, USA} \\
  \centerline{\normalfont $^4$Cedars-Sinai Medical Center, USA} \\[1ex]
  \centerline{\small\texttt{\{pav.poulos, tstafylakis\}@aueb.gr, \{ksadorf, costas\}@caltech.edu}} \\
  \centerline{\small\texttt{costas.anastassiou@csmc.edu}}
}

\begin{document}

\maketitle

\lhead{Preprint. Under review.} %
\renewcommand{\headrulewidth}{0pt}

\begin{abstract}

Spiking Neural Networks (SNNs) promise energy-efficient inference
through sparse, event-driven computation, yet remain constrained on
continuous, context-dependent tasks such as automatic speech recognition
(ASR). A key bottleneck is the Leaky Integrate-and-Fire (LIF) neuron,
whose membrane decay erases information across pauses and phonetic
boundaries while it must both extract features and retain long-range
context. Inspired by cortical separation of nonlinear filtering and
somatic integration, we introduce the Spiking Ternary GRU (stGRU): a
ternary-weight gated recurrent unit performs temporal filtering, while
downstream Resonate-and-Fire (RF) neurons generate the binary spike stream 
driving the next block's event pathway. This makes binary event transmission a
native training constraint rather than a separate deployment-time
activation quantization step; together with ternary weights, learned
projections avoid dense floating-point matrix MACs. Trained end-to-end
on LibriSpeech-960, Small, Medium, and Large stGRUs reach $8.09\%$,
$7.04\%$, and $6.44 \pm 0.17\%$ WER on test-clean without an external
language model. To our knowledge, these are the first matrix-MAC-free
spiking networks without a learned ANN sequence decoder to achieve
single-digit WER on LibriSpeech. The architecture also achieves
$10.88\%$ test CER on Aishell-1 ($9.28\%$ on dev) when trained from
scratch. Internally, trained networks develop a depth-graded hierarchy:
shallow layers act as fast acoustic filters, while deep layers form
slower semantic integrators whose latent representations segregate into
part-of-speech manifolds and word-level spike events. Deep-layer firing
rates follow lognormal distributions with a sparsity gradient, mirroring
hierarchical processing observed in human cortex in vivo. Our
biologically grounded architecture combines competitive ASR with
favorable modeled compute and memory cost while exposing interpretable
internal dynamics.
\end{abstract}

\section{Introduction}
\label{sec:introduction}

\begin{figure}[!t]
    \centering
    \vspace{-0.6\baselineskip}
    \begin{subfigure}[b]{\linewidth}
        \centering
        \includegraphics[width=\linewidth]{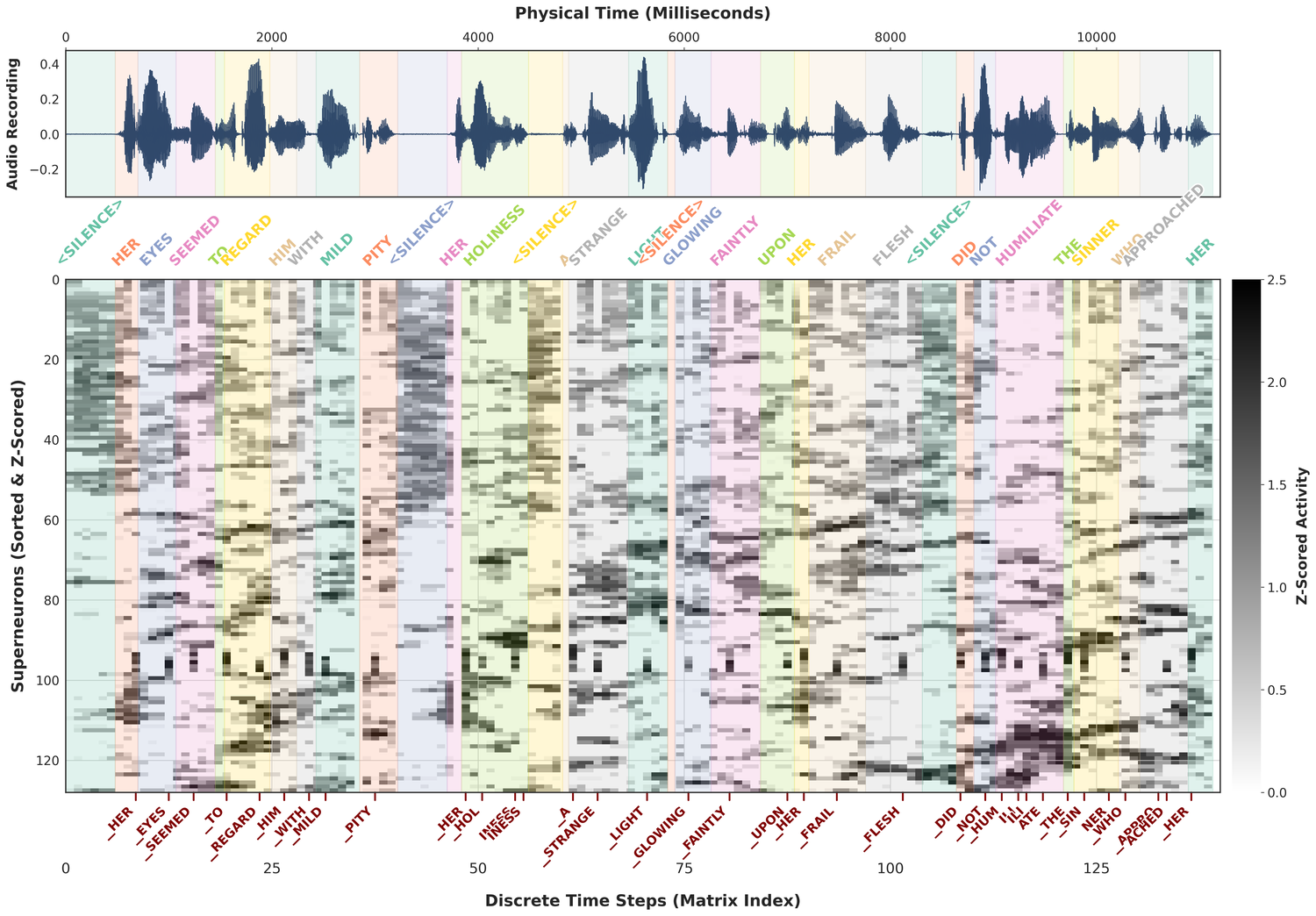}
        \label{fig:raster_plots}
    \end{subfigure}
    \vspace{-0.5\baselineskip}
    \begin{subfigure}[b]{\linewidth}
        \centering
        \includegraphics[width=\linewidth]{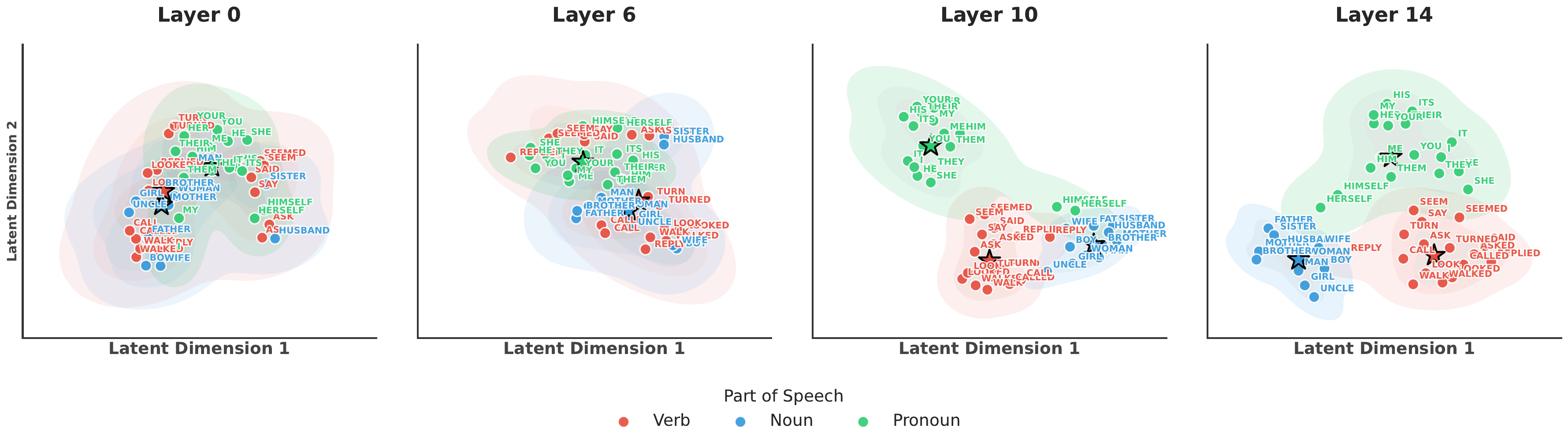}
        \label{fig:alignment_and_manifold}
    \end{subfigure}
    \captionsetup{skip=2pt}
    \caption{\textbf{Acoustic alignment, internal population activity, and semantic manifold emergence in stGRU.} 
    \textbf{(Top)} Input acoustic waveform (LibriSpeech test utterance) aligned with ground-truth word boundaries (color-shaded spans) and internal spiking representations across physical time (top axis) and discrete network timesteps (bottom axis). The central heatmap displays the $Z$-scored activity of 128 superneurons (co-active neural clusters sorted via Rastermap~\cite{rastermap2025}), with darker pixels indicating higher population firing. Red labels beneath the raster mark subword BPE token emissions predicted by the CTC readout layer (leading underscores denote new word onsets; consecutive identical tokens are collapsed into a single emission; CTC blanks omitted for clarity). 
    \textbf{(Bottom)} 2D UMAP projections of latent representations across network depth (layers 0, 6, 10, and 14). Internal states progressively disentangle into distinct semantic and syntactic clusters, color-coded by part of speech: verbs (red), nouns (blue), and pronouns (green).}
    \label{fig:overall_placeholder}
    \vspace{-1.2\baselineskip}
\end{figure}

\begin{figure}[!t]
    \centering
    \vspace{-0.6\baselineskip}
    
    \begin{subfigure}[b]{\linewidth}
        \centering
        \includegraphics[width=\linewidth]{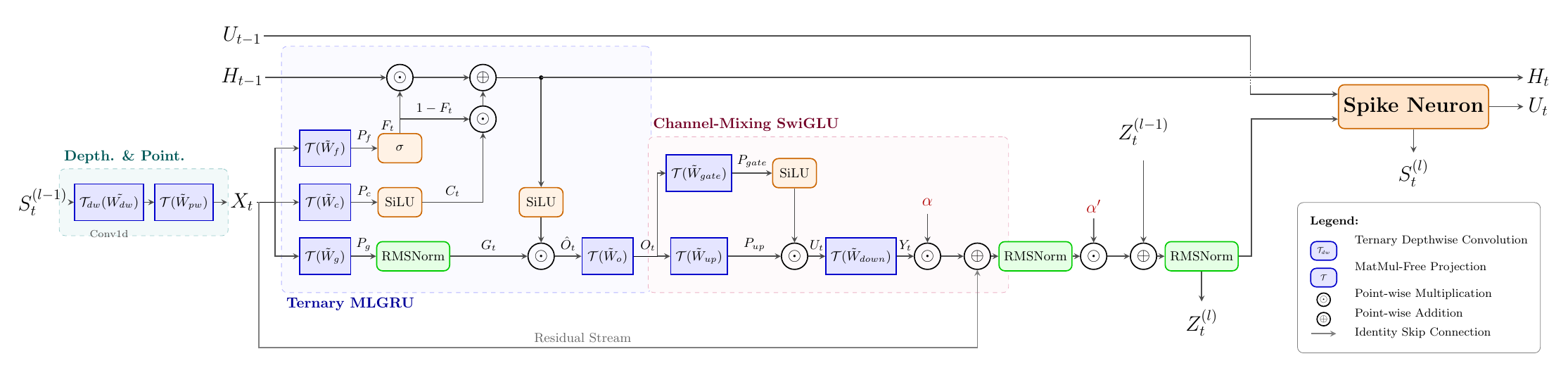}
        \caption{Proposed Architecture}
        \label{fig:proposed_architecture} %
    \end{subfigure}
    
    \vspace{0.3\baselineskip}   %
    
    \begin{subfigure}[b]{0.48\linewidth}
        \centering
        \includegraphics[width=\linewidth]{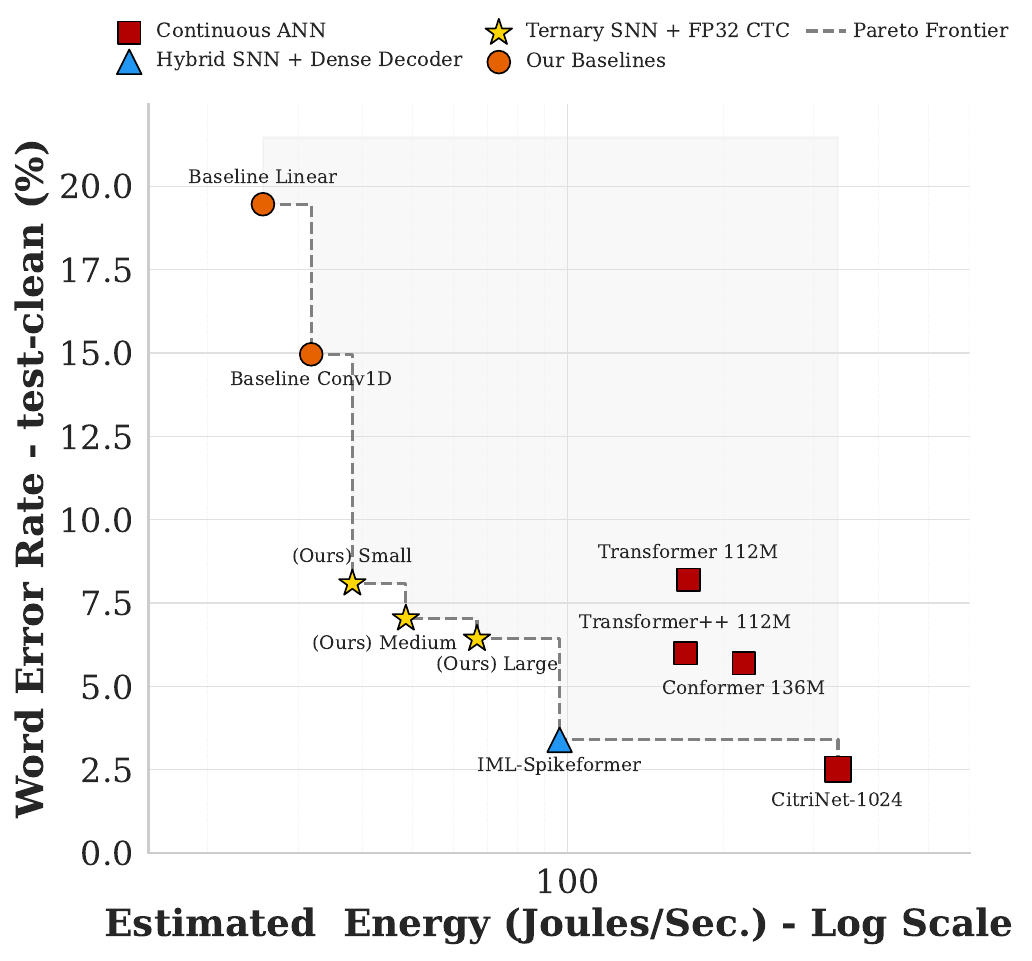}
        \caption{Energy Efficiency Frontier}
        \label{fig:pareto_energy}
    \end{subfigure}
    \hfill
    \begin{subfigure}[b]{0.48\linewidth}
        \centering
        \includegraphics[width=\linewidth]{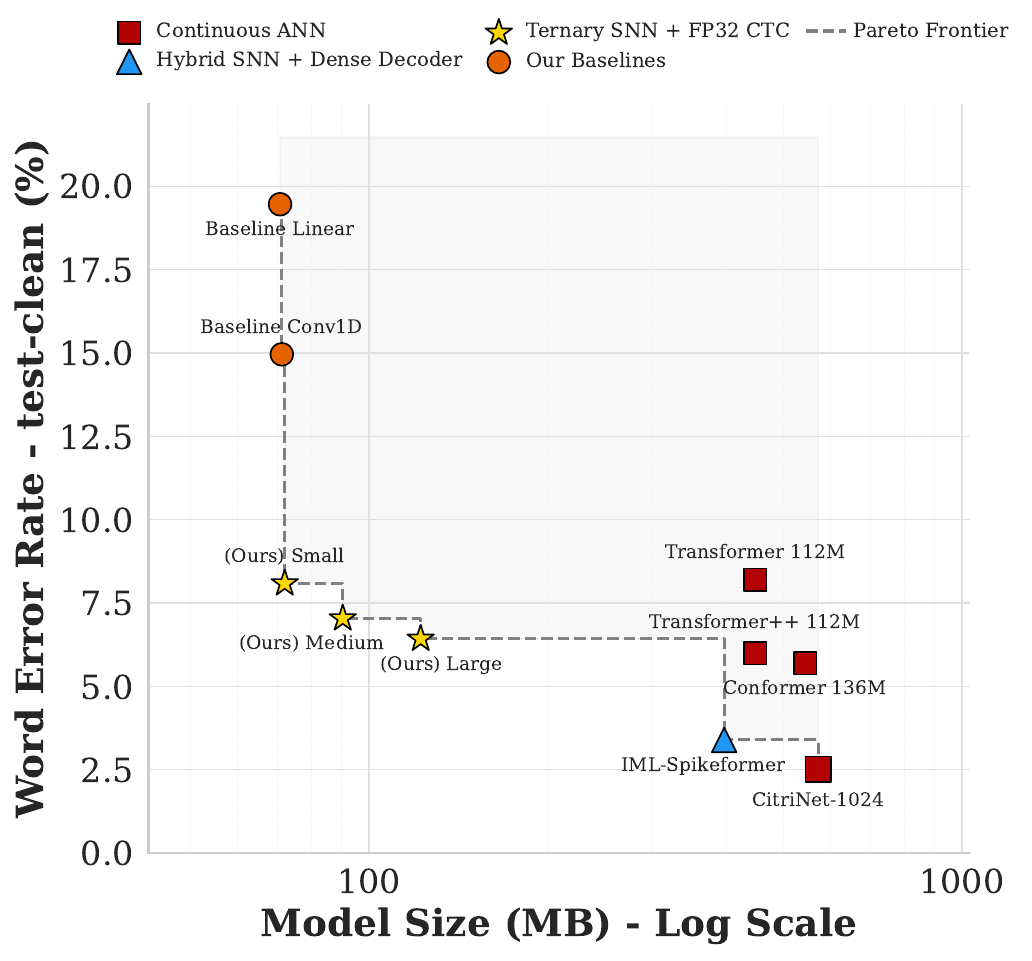}
        \caption{Memory Efficiency Frontier}
        \label{fig:pareto_memory}
    \end{subfigure}
    
    \captionsetup{skip=2pt}
    \caption{\textbf{Architecture and efficiency comparisons.}
\textbf{(a)} stGRU block architecture.
\textbf{(b)} LibriSpeech test-clean WER versus estimated arithmetic
energy over the 19,440\,s evaluation set.
\textbf{(c)} WER versus theoretical parameter storage in decimal MB.
The Large stGRU WER is a six-seed mean; energy estimates correspond
to the reported profiling run.}
    \label{fig:combined_results_and_pareto}
    \vspace{-1.2\baselineskip}
\end{figure}

Spiking Neural Networks (SNNs) promise highly energy-efficient inference by leveraging sparse, event-driven computations \cite{towards_roy_2019}. Yet they remain fundamentally constrained on continuous, context-dependent tasks such as automatic speech recognition (ASR) \cite{surrogate_bittar_2022, imlspikeformer_song_2025, deep_wu_2020}. 
The primary bottleneck lies in the standard Leaky Integrate-and-Fire (LIF) neuron \cite{spiking_gerstner_2002, networks_maaswofgang_1997}, whose intrinsic membrane decay erases critical context across pauses and phonetic boundaries \cite{long_bellec_2018, superspike_zenke_2018}.
Consequently, while continuous ANNs achieve state-of-the-art ASR performance \cite{gulati2020conformer, vaswani2017attention}, their reliance on dense multiply-accumulate (MAC) operations incurs prohibitive memory and energy costs \cite{11_horowitz_2014}-leaving a critical gap for energy-constrained applications.

Prior spiking ASR systems largely resorted to hybrid architectures pairing a spiking encoder with a dense, floating-point ANN sequence decoder, forfeiting the energy benefits of a truly edge-ready system \cite{surrogate_bittar_2022, imlspikeformer_song_2025, deep_wu_2020}. Deep event-driven architectures are further hindered by an $\mathcal{O}(L^{2})$ training bottleneck arising from sequential recurrent dependencies \cite{surrogate_neftci_2019, backpropagation_werbos_1990} and discontinuous spike resets \cite{training_lee_2016, surrogate_neftci_2019}. Together, these limitations motivate separating rich temporal filtering
from discrete event communication while keeping learned projections
matrix-MAC-free.

A complementary limitation appears in MatMul-free recurrent models:
ternary weights remove dense learned matrix products, but the MLGRU
backbone retains continuous activations, so deployment on low-precision
neuromorphic hardware requires a separate activation-quantization stage
\cite{scalable_zhu_2024}. This motivates our central question: can the
activation constraint itself be learned while retaining gated
long-range temporal filtering?

We answer this by treating spiking as a trainable activation interface
for the event-driven pathway. In stGRU, a ternary MLGRU performs
nonlinear temporal filtering, while a downstream RF neuron emits the
binary spike stream that drives the next block's event pathway.
In the ternary backbone, weights discretize what the network stores;
RF spikes discretize what its event pathway communicates. Crucially, the
spike constraint is present during optimization rather than imposed only
at deployment. This separation also mirrors cortical neurons, where
gated nonlinear processing precedes somatic integration 
\cite{gidon2020, dendrites_husser_2003, dendritic_london_2005,
mel1991, mel1992, shai2014, shai2015}. This storage--communication duality is central to stGRU.
Our primary contributions are as follows:

\textbf{Train-Time Dual Discretization \& matrix-MAC-free ASR.}

stGRU pairs ternary learned weights with an RF-generated binary event
stream, moving activation discretization into training rather than
leaving it solely to deployment. To our knowledge, it is the first
matrix-MAC-free SNN without a learned ANN sequence decoder to achieve
single-digit WER on LibriSpeech \cite{librispeech_panayotov_2015}:
Small, Medium, and Large stGRUs reach 8.09\%, 7.04\%, and
$6.44 \pm 0.17\%$ on test-clean without an external LM. Training from scratch on Aishell-1 \cite{aishell1_bu_2017} yields 10.88\%
test CER (9.28\% dev), close to the VGG-BiLSTM baseline (10.7\% test CER)
but below the strongest dense and hybrid systems in Table~\ref{tab:comprehensive_aishell}, providing cross-lingual evidence that the architecture transfers from English to Mandarin ASR.

\textbf{Exceptional Performance-to-Model-Size Trade-off.} 
Operating at $1.95\text{ mJ/s}$ (Small) and $3.42\text{ mJ/s}$ (Large), stGRU operates far below continuous Transformers ($8.80\text{ mJ/s}$) and Conformers ($11.28\text{ mJ/s}$) \cite{gulati2020conformer, 11_horowitz_2014}. For the Large Conv1D--MLGRU--ReLU baseline, ternarizing the weights changes
test-clean WER from 5.83\% to 6.49\%, an increase of 0.66 percentage points,
while the modeled arithmetic-energy rate changes from 33.48 to
4.10\,mJ/s (seed 3407). Ternary INT2 weights reduce the Large models to 122.22\,MB on
LibriSpeech and 167.84\,MB on Aishell-1, yielding a favorable modeled
energy--storage trade-off.

\textbf{Emergent ACC-like Error-Monitoring.} 
Layer L11 autonomously develops an ACC-like error-monitoring signature \cite{alexanderbrown2018, bryden2019}: incorrect words trigger elevated firing, peak sparsity (0.83), and heavy-tailed lognormal rate statistics ($\Delta R^2 = +0.091$ over Gaussian).

\textbf{Decision Confidence \& Autonomous ACC$\rightarrow$PFC Handoff.} 
At L14, the coding regime inverts into a prefrontal cortex (PFC)-like commitment state \cite{kianishadlen2009, rutishauser2015}: high-confidence correct words elicit peak firing, sparsity drops, and distributions shift to Gaussian. Together, these signatures suggest an autonomous ACC$\rightarrow$PFC-like functional handoff \cite{alexanderbrown2018} emerging within a single artificial sequence model.

\section{Related Work}
\label{sec:related_work}
While dense continuous models (Transformers, Conformers, CitriNet) define state-of-the-art ASR \cite{revisiting_hou_2024, citrinet_majumdar_2021, robust_radford_2022}, their reliance on dense multiply-accumulate (MAC) operations incurs prohibitive compute and memory costs \cite{11_horowitz_2014}. MatMul-free MLGRU combines ternary weights with continuous activations;
its Loihi~2 mapping therefore separately quantizes activations and
retains 16-bit precision after 8-bit quantization substantially degrades
accuracy \cite{scalable_zhu_2024}.

Motivated by biological efficiency, Spiking Neural Networks (SNNs) have been adapted for speech via surrogate gradients \cite{surrogate_bittar_2022, deep_wu_2020}, TCNs \cite{empirical_bai_2018}, and Spike-driven Transformers \cite{spikedriven_yao_2023, spikedriven_yao_2024}. However, existing ternary spiking speech models remain heavily restricted: QT-SNN \cite{qtsnn2025} targets only isolated keyword spotting, whereas DyTr-SNN \cite{dytrsnn2023} is evaluated solely on single-speaker LJSpeech without reporting WER and relies on a dense continuous 6-layer Transformer decoder. More broadly, leading SNN ASR baselines (GSU \cite{towards_hao_2024}, Spike-driven Transformer \cite{spikedriven_yao_2024}, IML-Spikeformer \cite{imlspikeformer_song_2025}) depend on continuous 6-layer ANN decoders to maintain competitive accuracy \cite{imlspikeformer_song_2025}. While their encoders use sparse accumulate (AC) operations, continuous decoders reintroduce a floating-point bottleneck and serialized execution, compromising edge viability.

stGRU bridges these two lines of work: it retains ternary MLGRU
filtering but trains an RF-generated binary event interface end-to-end,
making event discretization part of the model rather than a post-training
hardware conversion. Deferred scaling keeps learned backbone projections
matrix-MAC-free, while direct CTC decoding avoids a learned ANN sequence
decoder.

\section{Methodology}
\label{sec:methodology}

\subsection{Ternary Projections with Deferred Weight Scaling}
\label{sec:ternary_operators}

We use ternary BitLinear projections
\cite{era_ma_2024, scalable_zhu_2024}. For a full-precision
shadow-weight matrix $W \in \mathbb{R}^{m \times d}$, let
$\gamma_W = (md)^{-1}\sum_{i,j}|W_{ij}|$ and quantize by
\begin{equation}
    \widetilde W =
    \operatorname{RoundClip}
    \left(\frac{W}{\gamma_W+\epsilon},-1,1\right),
    \qquad
    \widetilde W_{ij}\in\{-1,0,+1\}.
\end{equation}
The shadow weights are trained with a straight-through estimator
(STE) \cite{estimating_bengio_2013} and QAT warmup
(Appendix~\ref{app:qat_warmup}).

Let $V$ denote the input presented to a projection after its shared
input normalization. The three ablation modes differ only in the
placement or omission of the weight scale:
\begin{equation}
\begin{aligned}
    Y_{\mathrm{naive}}
        &= V(\gamma_W\widetilde W)^\top,\\
    Y_{\mathrm{proposed}}
        &= (V\widetilde W^\top)\gamma_W,\\
    Y_{\mathrm{none}}
        &= V\widetilde W^\top.
\end{aligned}
\label{eq:scaling_modes}
\end{equation}
The naive and proposed modes are algebraically equivalent for fixed
$V$, $\widetilde W$, and $\gamma_W$. Naive scaling folds $\gamma_W$
into the weights, producing a floating-point weight matrix.
Deferred scaling instead accumulates through ternary weights using
additions and subtractions, then applies $\gamma_W$ to the output.
The None mode omits this scale entirely. Thus the proposed mode
avoids dense floating-point matrix MACs in the learned projection,
but retains an output scaling operation. We write this operator as
\begin{equation}
    \mathcal T(V,\widetilde W,\gamma_W)_i
       = \gamma_W\sum_{j=1}^{d}\widetilde W_{ij}V_j,
       \qquad i=1,\ldots,m.
    \label{eq:t_operator}
\end{equation}

The same scale placement applies to a temporal depthwise projection.
Let $V\in\mathbb R^{T\times C}$ be the input to that projection and
$\widetilde W_{dw}\in\{-1,0,+1\}^{K\times C}$ its ternary kernel.
With a weight scale $\gamma_{dw}$, we define
\begin{equation}
    \mathcal T_{dw}(V,\widetilde W_{dw},\gamma_{dw})_{t,c}
      = \gamma_{dw}
        \sum_{k}\widetilde W_{dw,k,c}V_{t+k,c}.
    \label{eq:dw_conv}
\end{equation}
The temporal sum therefore uses additions and subtractions through
ternary weights; $\gamma_{dw}$ is applied after that sum. This avoids folding a floating-point scale into every kernel weight.
In the ablation (Table~\ref{tab:ablation_deferred_scaling}),
deferred scaling reduces the modeled energy rate by about
$5\times$ relative to naive scaling, with comparable test-clean WER
(Small: 8.09\% vs.\ 8.15\%; Large: 6.14\% vs.\ 6.44\%).
Omitting the scale instead worsens test-clean WER to 9.44\%
and 7.51\%, respectively.

\subsection{Parallel Spiking-Ternary Block Architecture}

To bypass the prohibitive $\mathcal{O}(L^2)$ training bottleneck of recurrent networks and discontinuous resets \cite{backpropagation_werbos_1990}, the Spiking-Ternary GRU Block decouples recurrent filtering from spike emission, enabling global parallel unrolling in $\mathcal{O}(L \log L)$ time while avoiding dense learned matrix MACs at inference.

Acoustic features are initially mapped by a pre-encoder into discrete spike trains $S^{(0)}_t \in \{0, 1\}^{C}$. At block $l$, temporal features are extracted from the incoming discrete spike sequence $S^{(l)} \in \{0, 1\}^{T \times C}$ via a depthwise-separable operator pair consisting of a 1D temporal depthwise convolution and a pointwise channel projection:
\begin{equation}
    X_{dw} = \mathcal{T}_{dw}(S^{(l)}, \widetilde{W}_{dw}, \gamma_{dw}), \quad X = \mathcal{T}(X_{dw}, \widetilde{W}_{pw}, \gamma_{pw})
\end{equation}
where $\widetilde{W}_{pw} \in \{-1, 0, +1\}^{C \times C}$ denotes the ternary pointwise weights and $\gamma_{pw}$ their scale.

We adapt the MLGRU of \citet{scalable_zhu_2024} with deferred-scaling
ternary projections, RMS-normalized gating, and a ternary output projection. For brevity, we denote gate-specific ternary projections as $\mathcal{T}_k(V) \triangleq \mathcal{T}(V, \widetilde{W}_k, \gamma_k)$ for $k \in \{c, f, g\}$. For input representation $X_t \in \mathbb{R}^C$, the gates are computed as $C_t = \text{SiLU}(\mathcal{T}_c(X_t))$, $F_t = \sigma(\mathcal{T}_f(X_t))$, and $G_t = \text{RMSNorm}(\mathcal{T}_g(X_t))$, updating state and output via:
\begin{align}
    H_t &= F_t \odot H_{t-1} + C_t \odot (1 - F_t) \label{eq:mlgru_h} \\
    O_t &= \mathcal{T}(G_t \odot \text{SiLU}(H_t), \widetilde{W}_o, \gamma_o)
\end{align}
Because Equation~\ref{eq:mlgru_h} is strictly element-wise (no dense $Wh_{t-1}$ products), intermediate representations are computed globally across sequence length $T$ via an associative parallel scan \cite{efficiently_gu_2021, simplified_smith_2022}.

Channels are mixed via a ternary SwiGLU block stabilized by learnable ReZero vectors $\alpha \in \mathbb{R}^C$ \cite{rezero_bachlechner_2021} (Appendix \ref{app:rezero}):
\begin{align}
    U_t &= \text{SiLU}(\mathcal{T}(O_t, \widetilde{W}_{\text{gate}}, \gamma_{\text{gate}})) \odot \mathcal{T}(O_t, \widetilde{W}_{\text{up}}, \gamma_{\text{up}}) \\
    Z_t &= X_t + \alpha \odot \mathcal{T}(U_t, \widetilde{W}_{\text{down}}, \gamma_{\text{down}}) \label{eq:current_z}
\end{align}

To prevent vanishing gradients and stabilize variance across depth, the continuous residual stream $Z^{(l)} \in \mathbb{R}^C$ updates via:
\begin{equation}
    Z^{(l)}_t = \text{RMSNorm}(Z^{(l-1)}_t) + \alpha^{'} \odot \text{RMSNorm}(Z_t)
\end{equation}

The stabilized residual $Z^{(l)}_t$ serves as the input current to the Parallel Resonate-and-Fire (PRF) neuron \cite{prf_huang_2024}. Unlike standard LIF neurons which enforce sequential BPTT due to hard reset subtractions ($V_{th} S_{t-1}$), PRF folds resets into a complex decay factor $\tilde{\gamma} = -\frac{1}{\tau} + i\theta$. Under Euler discretization over timestep $\Delta$, membrane integration unfolds linearly:
\begin{equation}
    U_t = \exp\left(-\frac{\Delta}{\tau}\right)\left(\cos(\Delta\theta) + i\sin(\Delta\theta)\right)U_{t-1} + \Delta Z^{(l)}_t \label{eq:prf_euler}
\end{equation}
Because temporal integration is independent of output spikes, membrane potentials $U_{1:T}$ evaluate globally in the frequency domain via Fast Fourier Transform ($\mathcal{F}$) \cite{algorithm_cooley_1965, efficiently_gu_2021} using decay kernel $K_{1:T} = (\Delta A^0, \dots, \Delta A^{T-1})$ with $A = \exp(\Delta \tilde{\gamma})$:
\begin{equation}
    U_{1:T} = \mathcal{F}^{-1}(\mathcal{F}(Z^{(l)}_{1:T}) \odot \mathcal{F}(K_{1:T})) \label{eq:prf_fft}
\end{equation}
Finally, the continuous potentials are folded back into the binary domain for the $(l+1)$-th block by applying a Heaviside step function strictly to the real component:
\begin{equation}
    S^{(l+1)}_t = \mathcal{H} (\Re\{U_t\}- V_{th} ) \label{eq:prf_spike}
\end{equation}

Surrogate gradients (Multi-Gaussian \cite{accurate_yin_2021}) train
this binary event interface end-to-end, so the next block is optimized
on the same discrete spike stream used at inference. The MLGRU recurrence
admits an associative scan and PRF integration an
$\mathcal{O}(L\log L)$ FFT implementation; empirically, training is
3.1$\times$--3.4$\times$ faster than the parameter-matched LIF variants
(Tables~\ref{tab:ablation_study}, \ref{tab:lif_vs_prf_time}).

\begin{table}[t]
\centering
\caption{Comprehensive Architectural Comparison on LibriSpeech-960 ASR. All reported WER results evaluate the standalone acoustic model without an external language model. The Large RF WER is mean $\pm$ standard deviation over six seeds; other
in-house rows and arithmetic-energy rates correspond to seed 3407.}
\label{tab:comprehensive_asr}
\resizebox{\textwidth}{!}{
\begin{tabular}{lrlcccc}
\toprule
\multirow{2}{*}{\textbf{Architecture}} & \multirow{2}{*}{\textbf{Params (M)}} & \multirow{2}{*}{\textbf{Weight Type}} & \textbf{Memory} & \textbf{Energy rate (mJ/s)} & \multicolumn{2}{c}{\textbf{WER (\%)}} \\ \cmidrule(lr){6-7} 
 & & & \textbf{(MB)} & \textbf{(mJ/s)} & \textbf{test-clean} & \textbf{test-other} \\ 
\midrule
\multicolumn{7}{c}{\textit{Continuous Artificial Neural Networks (ANNs)}} \\
\midrule
Moritz et al. \cite{triggered_moritz_2019}    & 70.0 & FP32 & $280.0^\dagger$ & - & $5.4$ & $16.1$ \\
Park et al. \cite{specaugment_park_2019}      & 360.0 & FP32  & $1440.0^\dagger$ & - & $2.8$ & $6.8$ \\
Whisper-Small \cite{robust_radford_2022}      & 244.0 & FP16  & $488.0^\dagger$ & - & $3.1$ & $7.4$  \\
ResNet-BiGRU \cite{customized_reza_2022}      & 33.0 & FP32  & $132.8^\dagger$ & - & $15.6$ & - \\
Quantized RNN-T \cite{sub8bit_zhen_2022}      & 27.8 & FP8  & $27.8^\dagger$ & - & $8.70$ & $22.36$ \\
Quantized RNN-T \cite{sub8bit_zhen_2022}      & 27.8 & FP4  & $13.9^\dagger$ & - & $8.78$ & $22.54$ \\
Transformer $^*$ \cite{revisiting_hou_2024}   & 112.0 & FP32  & $448.0^\dagger$ & 8.80$^{\dagger\dagger}$ & $8.2$ & $15.2$ \\
Transformer++ $^*$ \cite{revisiting_hou_2024} & 112.0 & FP32  & $448.0^\dagger$ & 8.69$^{\dagger\dagger}$ & $6.0$ & $11.8$ \\
Conformer  $^*$ \cite{revisiting_hou_2024}    & 136.0 & FP32  & $544.0^\dagger$ & 11.28$^{\dagger\dagger}$ & $5.7$ & $10.8$ \\
CitriNet-1024 \cite{citrinet_majumdar_2021}     & 143.1 & FP32  & $572.4^\dagger$ & 17.18$^{\dagger\dagger}$ & $2.52$ & $6.22$ \\

\midrule
\multicolumn{7}{c}{\textit{Hybrid Spiking Neural Networks (SNNs) \& Dense ANN Decoder/LM}} \\
\midrule
Deep FC SNN (LS-360h) \cite{deep_wu_2020} & - & FP32  & -  & - & $9.4$ & - \\
1 Spiking RNN + 3 LSTM \cite{surrogate_bittar_2022} & - & FP32 & - & - & $3.51$ & - \\
4 Spiking RNN + 0 LSTM \cite{surrogate_bittar_2022} & - & FP32  & - & - & $9.94$ & - \\
GSU (+ 6-layer ANN Dec) $^{**}$ \cite{towards_hao_2024} & 185.0 & FP32 & $740.4^\dagger$ & - & $12.8$ & $32.1$ \\
Spike-driven Trans. (+ 6-layer ANN Dec) $^{**}$ \cite{spikedriven_yao_2024} & 99.0 & FP32& $397.2^\dagger$ & - & $10.5$ & $25.7$ \\
IML-Spikeformer (+ 6-layer ANN Dec) \cite{imlspikeformer_song_2025} & 99.0 & FP32 & $397.2^\dagger$ & 4.96$^{\dagger\dagger}$ & \textbf{3.4}  & \textbf{7.9}  \\

\midrule
\multicolumn{7}{c}{\textit{Ours (Structural Baselines, FP32 \& W/O Learned ANN Sequence Decoder)}} \\
\midrule
\textbf{Large} (Linear $\rightarrow$ RF) & 77.01 & FP32$^\ddagger$ & 117.28 & 1.30 & 17.34 & 33.64 \\
\textbf{Large} (Conv1D $\rightarrow$ RF) & 77.58 & FP32$^\ddagger$ & 119.56 & 2.22 & 11.64 & 26.20 \\

\midrule
\multicolumn{7}{c}{\textit{Ours (Parameter-Matched Baselines, Continuous Dense ReLU Activations)}} \\
\midrule
Small (Conv1D $\rightarrow$ MLGRU $\rightarrow$ ReLU)   & 132.66 & FP32  & 530.63  & 22.12 & 7.25 & 19.14 \\
Medium (Conv1D $\rightarrow$ MLGRU $\rightarrow$ ReLU)  & 204.16 & FP32  & 816.63  & 26.25 & 6.41 & 17.29 \\
Large (Conv1D $\rightarrow$ MLGRU $\rightarrow$ ReLU)   & 329.39 & FP32  & 1317.55 & 33.48 & 5.83 & 16.11 \\
Small (Conv1D $\rightarrow$ MLGRU $\rightarrow$ ReLU)   & 132.66 & Ternary INT2  & 72.01  & 2.60  & 7.74 & 20.26 \\
Medium (Conv1D $\rightarrow$ MLGRU $\rightarrow$ ReLU) & 204.16 & Ternary INT2  & 90.16   & 3.13  & 6.68 & 17.71 \\
Large (Conv1D $\rightarrow$ MLGRU $\rightarrow$ ReLU)   & 329.39 & Ternary INT2  & 121.96  & 4.10  & 6.49 & 17.40 \\

\midrule
\multicolumn{7}{c}{\textit{Ours (matrix-MAC-free, Ternary Spiking Networks \& W/O Learned ANN Sequence Decoder})} \\
\midrule
\textbf{Ours Large (Streaming) (Conv1D $\rightarrow$ MLGRU $\rightarrow$ RF)} & 329.45 & Ternary INT2 & 122.22 & 3.51 & 9.11 & 22.13 \\
\textbf{Ours Small} (Conv1D $\rightarrow$ MLGRU $\rightarrow$ RF)  & 132.68 & Ternary INT2 & 72.09 & 1.95 & 8.09 & 20.79 \\
\textbf{Ours Medium} (Conv1D $\rightarrow$ MLGRU $\rightarrow$ RF) & 204.20 & Ternary INT2 & 90.31 & 2.49 & 7.04 & 18.74 \\
\textbf{Ours Large} (Conv1D $\rightarrow$ MLGRU $\rightarrow$ RF)  & 329.45 & Ternary INT2 & 122.22 & 3.42 & $6.44 \pm 0.17$ & $17.49 \pm 0.23$ \\
\bottomrule
\multicolumn{7}{l}{\scriptsize $^\dagger$ Theoretical memory size computed based on parameter count and format precision (FP32 = 4 Bytes, FP16 = 2 Bytes, INT2 = 0.25 Bytes).} \\
\multicolumn{7}{l}{\scriptsize $^{\dagger\dagger}$ Our Reproduced results by publicly available data.} \\
\multicolumn{7}{l}{\scriptsize $^*$ Trained on 2.5k hours of various datasets as cited in \cite{revisiting_hou_2024}.} \\
\multicolumn{7}{l}{\scriptsize $^{**}$ Reproduced by \cite{imlspikeformer_song_2025}.} \\
\multicolumn{7}{l}{\scriptsize $^\ddagger$ Structural baselines use an
INT2 ternary downsampler and FP32 downstream weights; unmarked FP32
variants also use an FP32 downsampler.} \\
\end{tabular}
}
\vspace{-0.3cm}
\end{table}

\section{Experiments}
\label{sec:experiments}

We evaluate the Spiking Ternary GRU (stGRU) on LibriSpeech-960 and Aishell-1 \cite{librispeech_panayotov_2015, aishell1_bu_2017} implemented in SpeechBrain \cite{speechbrain_ravanelli_2022} across 8 AMD Instinct MI250X GCDs on the LUMI supercomputer. Acoustic inputs (80-dimensional log-Mel filterbanks at 50\,Hz) undergo $4\times$ temporal subsampling via a ternary 2D CNN frontend (512 and 1024 channels), followed by an MLGRU layer coupled with a Parallel Resonate-and-Fire (PRF) neuron acting as a pre-encoder that emits discrete spikes at a reduced rate of 12.5\,Hz. The subsequent stGRU backbone applies 1D depthwise convolutions with kernel sizes up to $K=63$ across 1024 channels (Appendix \ref{app:architecture}), trained end-to-end via CTC loss \cite{connectionist_graves_2006} with Semi-Resonant Cosine Annealing (Appendix \ref{app:lr_scheduler}). All primary metrics evaluate the standalone acoustic model; an optional OpenSLR 4-gram language model\footnote{\url{https://www.openslr.org/11/}} is decoded (Table \ref{tab:large_prf_seed_runs}) via non-neural CPU hash lookups and is therefore excluded from neural energy profiling.

\textbf{Performance and Efficiency Benchmarks.} 
We benchmark against continuous ANNs (Transformer, Conformer, CitriNet \cite{vaswani2017attention, revisiting_hou_2024, gulati2020conformer, citrinet_majumdar_2021}) and the hybrid SNN IML-Spikeformer \cite{imlspikeformer_song_2025}, calculating theoretical energy via 45\,nm CMOS constants (4.6\,pJ/FP-MAC, 0.9\,pJ/sparse-AC, 17.5\,pJ/PRF-update \cite{11_horowitz_2014, prf_huang_2024}).
As reported in Table~\ref{tab:comprehensive_asr}, Small, Medium, and Large stGRUs achieve standalone WERs of 8.09\%, 7.04\%, and $6.44 \pm 0.17\%$ on LibriSpeech test-clean without an external LM (optional 4-gram LM yields $5.95 \pm 0.16\%$; Table~\ref{tab:large_prf_seed_runs}), setting the first single-digit WERs for matrix-MAC-free SNNs without a learned ANN sequence decoder. A bounded-latency streaming variant with 320\,ms look-ahead attains 9.11\% WER at 3.51\,mJ/s (Table~\ref{tab:comprehensive_asr}). A separately trained stGRU achieves 10.88\% test CER on Aishell-1 and 9.28\% on dev (Table~\ref{tab:comprehensive_aishell}).

Parameter-matched ablations isolate individual design choices (Table~\ref{tab:comprehensive_asr}, Table~\ref{tab:ablation_study}): quantizing from FP32 to ternary weights cuts memory by $7.4\times$--$10.8\times$ with negligible loss ($5.83\% \to 6.49\%$). At the Large scale, discrete PRF spikes surpass continuous ternary ReLUs in both WER and energy ($6.14\%$ at 3.42\,mJ/s vs.\ $6.49\%$ at 4.10\,mJ/s; widening to 33.48\,mJ/s for FP32 ReLUs; Table~\ref{tab:ablation_study}). Under the same analytical 45\,nm arithmetic-cost model, our estimate
for IML-Spikeformer allocates 2.73\,mJ/s to its FP32 frontend and
1.68\,mJ/s to its ANN decoder, whereas the stGRU estimate includes
0.05\,mJ/s for active-weight accumulation in its FP32 CTC readout
(Table~\ref{tab:energy_profiling_full}). These are operation-based
estimates, not measured hardware energy. Upgrading from Linear $\to$ RF to the full ternary stack adds $<5$\,MB and $+2.12$\,mJ/s while reducing test-clean WER by 11.20\% percentage points for seed 3407 ($17.34\% \to 6.14\%$).

\begin{figure}[!t]
    \centering
    \vspace{-0.6\baselineskip}  
    \includegraphics[width=0.9\linewidth]{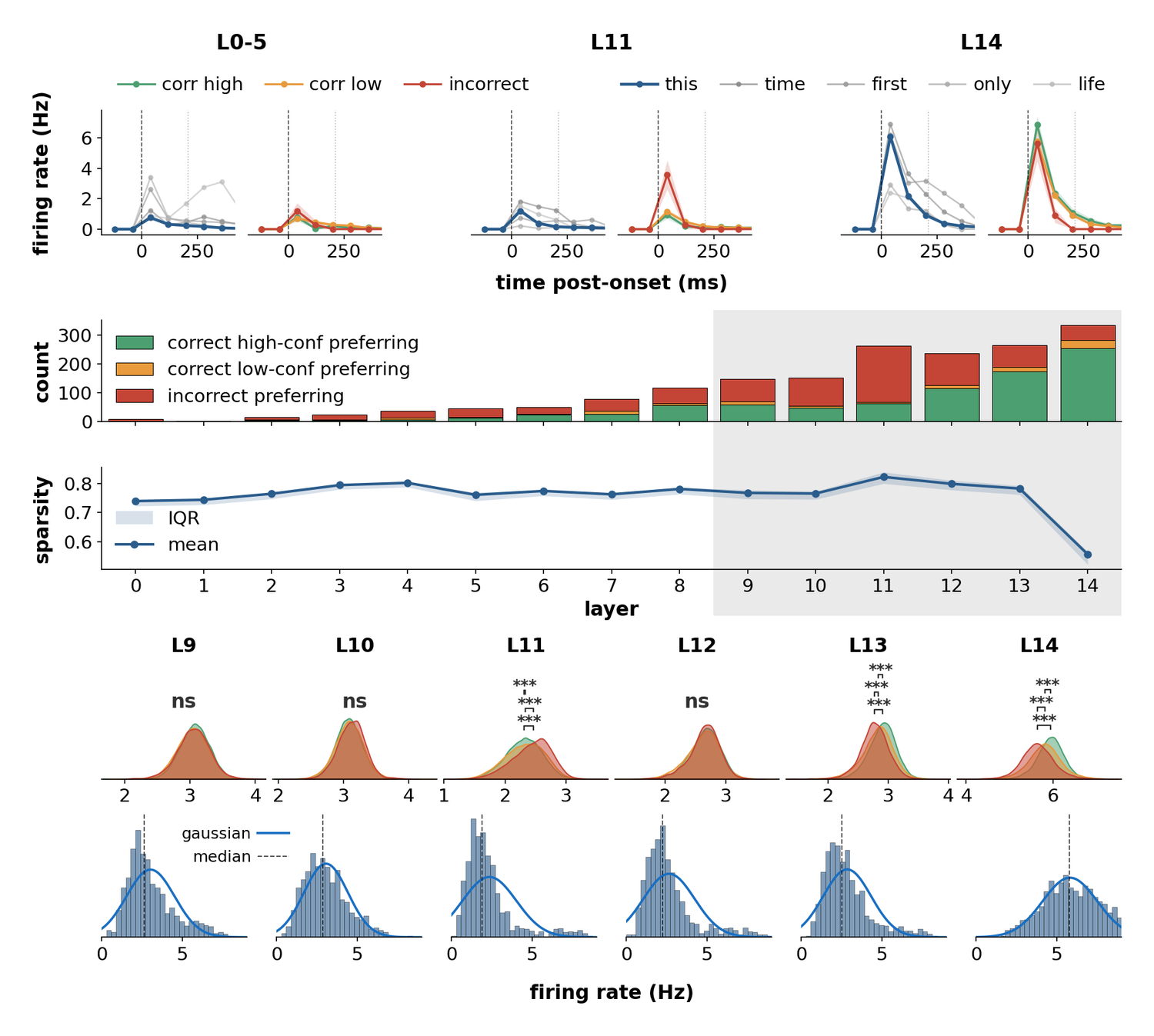}
    \captionsetup{skip=1pt} 
    \caption{\textbf{Depth-graded coding emerges in trained Spiking-Ternary networks.} Row 1: PSTH of word-selective neurons at shallow (L0--5), uncertainty (L11), and commitment (L14) depths. Each panel pair shows responses to a target word (``this,'' blue) against acoustically similar competitors (gray), with the right sub-panel split by outcome (green: high-conf correct; orange: low-conf correct; red: incorrect). Row 2: Per-layer counts of outcome-selective neurons (matched-AUC, FDR $q<0.001$); correct-preferring neurons are sub-classified by high- vs.~low-confidence via median CTC log-probability. Selectivity grows from L0--L5 to L11--L14 ($<50$ vs. $>250$ neurons), and the bipartite partition reverses across depth: incorrect-preferring neurons peak at L11 ($n=194$), high-confidence-correct-preferring at L14 ($n=254$). Row 3: Population sparsity (per-word silent fraction) across encoder depth, with shaded IQR. Row 4: Population firing rate distributions for L9--L14 split by outcome (brackets: pairwise Mann--Whitney U-test with Bonferroni correction; \texttt{ns}: panels with at least one non-significant comparison). Row 5: Per-neuron firing rate histograms for L9--L14 with Gaussian fit overlaid.}
    \label{fig:neuromorphic_outcome_signature}
    \vspace{-1.2\baselineskip} 
\end{figure}

\section{Neuromorphic Interpretation}
\label{sec:neuromorphic}

Having established favorable modeled efficiency on English and Mandarin ASR, we now examine whether the network’s internal representations exhibit functional analogies to cortical decision signals. Echoing the parallel, distributed processing of human auditory cortex \cite{hamilton2021, bhayagrossman2022}, deep layers (L10--L14) autonomously partition into functionally distinct subpopulations whose activity statistics mimic those of neurons in human cortex during perceptual decisions. We substantiate this below.

\subsection{Error-Monitoring Signature in L11 Mimics the Human Medial Frontal Cortex}
At L11, incorrect words result in significantly higher population firing than either high- or low-confidence correct words (Fig.~\ref{fig:neuromorphic_outcome_signature}, row 4) analogous to error-related activity associated with anterior cingulate cortex (ACC) \cite{alexanderbrown2018, bryden2019}. The error code is carried by a subpopulation of neurons: incorrect-preferring neurons outnumber high-confidence-correct neurons ($n=194$ vs.~$64$; Fig.~\ref{fig:neuromorphic_outcome_signature}, top). L11 also exhibits the sparsest spiking activity from all layers (Fig.~\ref{fig:neuromorphic_outcome_signature}, row 3), placing it in the recruitment-based regime characteristic of cortical error monitoring. The response is also temporally appropriate for an error signal: incorrect-preferring neurons exhibit a sharp post-onset transient that decays before the CTC emission window, on the timescale of error detection rather than sustained stimulus encoding. Per-neuron firing rates at L11 follow a lognormal distribution (Fig.~\ref{fig:supp_firing_rate_lognormal_vs_gaussian}) as observed in cortical recall populations. It follows that directionality, recruitment, timing, and distribution align with an ACC-like role of L11.

\subsection{Decision-Confidence Signature in L14 Mimics the Prefrontal Cortex}
L14, in turn, exhibits different behavior to L11. In L14 the highest spike rate corresponds to high-confidence correct words while low-confidence correct words are between the high-confidence and incorrect distributions. This graded confidence axis is absent at L11 and mimics the behavior observed in primate lateral intraparietal (LIP) and human medial temporal lobe (MTL) \cite{kianishadlen2009, rutishauser2015}. At the level of decision hierarchy, prefrontal cortex (PFC) is the area where confidence signals from upstream regions are converted into committed action. The composition of the selective population inverts in parallel: high-confidence-correct neurons dominate ($n=254$ vs.~$58$ incorrect-preferring; Fig.~\ref{fig:neuromorphic_outcome_signature}, top). PSTH responses broaden in turn, replacing the narrow error-only transient at L11 with large, sustained post-onset peaks across all three outcome conditions---the response shape expected of a layer broadcasting a committed token rather than gating a selective monitoring signal. Moreover, population sparsity decreases significantly from L11 to L14 and the per-neuron firing-rate changes from lognormal to near-Gaussian ($\Delta R^2 = -0.011$, the only negative value in the recall hierarchy; Fig.~\ref{fig:supp_firing_rate_lognormal_vs_gaussian}). The latter indicates broad recruitment rather than selective monitoring with L14 emulating human PFC behavior.

\subsection{Autonomous ACC-to-PFC Handoff Within a Single Network}
The transition between L11 and L14 proceeds gradually through L12--L13 in a functional handoff. L12 acts as the inflection layer where responses to all categories become similar (Fig.~\ref{fig:neuromorphic_outcome_signature}, bottom, \texttt{ns}), the population code's directionality reverses, and the spike rate lognormality gradually decays from L11 ($+0.091$) through L12 ($+0.064$) and L13 ($+0.041$) before vanishing at L14 ($-0.011$). Selectivity composition, sparsity, firing-rate directionality, and distribution shape move together along this trajectory, consistent with a continuous functional gradient. This structure mimics the ACC with its error and surprise signals \cite{alexanderbrown2018} while PFC implements the downstream control that converts those signals into committed decisions. Critically, this handoff is implemented \textit{within} a single recurrent network rather than \textit{across} brain regions: a network trained end-to-end on CTC, with no architectural prior favoring functional segregation, partitions its deep layers into error-monitoring and confidence-readout subpopulations whose responses, sparsity, and firing-rate statistics jointly track signatures observed in human cortex \textit{in vivo}. We note that this partition is not incidental to depth or spiking activity alone: in an ablated variant without MLGRU gating, deep-layer outcome selectivity collapses by 95.5\% once empty-emission trials are controlled for, versus 8.1\% in the full model (Supp.\ Fig.~\ref{fig:mlgru_comparison}). This ablation supports a role for MLGRU gating in the observed
outcome-selective activity. This mirrors cortical pyramidal neurons, where nonlinear dendritic filtering precedes somatic integration \cite{dendritic_london_2005, mel1992, poirazi2003, shai2015} and subthreshold resonance bridges temporal context across pauses \cite{long_bellec_2018, giraud2012}: upstream recurrent gating provides the temporal substrate necessary for downstream somatic integrators to develop decision-level outcome selectivity.

\paragraph{Limitations.}
Our energy estimates use analytical operation counts and 45\,nm CMOS
constants. QAT also retains full-precision shadow weights and their training
overhead: the forward weights become ternary after warmup, while the
straight-through estimator updates the continuous weights during
backpropagation. Furthermore, while our design eliminates all dense
$\mathcal{O}(C^2)$ floating-point matrix multiplications, the model is
not entirely multiplication-free: continuous element-wise multiplications
persist in recurrent gating, residual scaling, RMSNorm, and neuron
dynamics. These operations are element-wise rather than dense \(O(C^2)\) matrix multiplications, but deploying them on pure addition-only or integer-only edge silicon still requires specialized arithmetic units or approximations. Performance remains weaker on
\texttt{test-other} than on \texttt{test-clean}. Finally, mitigating
the capacity limits of 2-bit weights required wide channels and
temporal kernels of up to $K=63$, despite the low precision of their
weights.

\section{Conclusion}
\label{sec:conclusion}

We introduced stGRU, an ASR architecture combining deferred-scaling
ternary projections, gated recurrent filtering, and resonant spiking
integration. At inference, its learned projections avoid dense matrix
MACs, while the Large model reaches $6.44 \pm 0.17\%$ LibriSpeech
test-clean WER without an external language model and a separately
trained Aishell-1 model reaches 10.88\% test CER. The modeled
arithmetic-energy and parameter-storage results show a favorable
efficiency trade-off, subject to the accounting limits above.

Conceptually, stGRU couples complementary low-precision constraints in the event-driven backbone:
ternary learned projections constrain synaptic weights, while RF neurons make the inter-block event stream binary during training rather than relying on a separate deployment-time activation-quantization step. Spiking thus serves not only as
neuronal dynamics but as an end-to-end optimized binary communication bottleneck. Despite this
constraint, deep layers still develop error-to-confidence selectivity
and structured sparsity, suggesting that efficient event-driven
communication can coexist with rich internal organization.

\bibliography{bibliography}
\bibliographystyle{iclr2027_conference}

\appendix

\setcounter{table}{0}
\renewcommand{\thetable}{S\arabic{table}}
\setcounter{figure}{0}
\renewcommand{\thefigure}{S\arabic{figure}}
\setcounter{equation}{0}
\renewcommand{\theequation}{S\arabic{equation}}
\setcounter{algorithm}{0}
\renewcommand{\thealgorithm}{S\arabic{algorithm}}

\section{Architectural Configurations and Streaming Setup}
\label{app:architecture}

Table \ref{tab:architecture_details} details the layer-by-layer configurations for the Small, Medium, and Large variants of our proposed stGRU architecture. The network hierarchically scales its temporal receptive field, mapping sequences from basic phonetic structures up to extended sentence and paragraph-level contexts. Across all scales, the temporal convolutions utilize a fixed channel dimension of 1024, a stride of 1, and a dilation of 1. Capacity is scaled by uniformly increasing the depth within each linguistic stage, culminating in a 15-layer Large variant that introduces an exclusive fifth stage with a receptive field of $K=63$.

\begin{table}[h]
\centering
\caption{\textbf{Detailed architectural configurations for the Small, Medium, and Large stGRU variants.} The receptive field expands progressively across stages to capture hierarchical linguistic features.}
\label{tab:architecture_details}
\vspace{0.15cm}
\begin{tabular}{lcccccc}
\toprule
\multirow{2}{*}{\textbf{Linguistic Stage}} & \multirow{2}{*}{\textbf{Channels}} & \multirow{2}{*}{\textbf{Kernel Size}} & \multicolumn{3}{c}{\textbf{Number of Layers}} \\ \cmidrule(lr){4-6}
 & & & \textbf{Small} & \textbf{Medium} & \textbf{Large} \\ 
\midrule
Stage 1 (Phonemes)         & 1024 & 11 & 1 & 2 & 3 \\
Stage 2 (Syllables)        & 1024 & 21 & 1 & 2 & 3 \\
Stage 3 (Words)            & 1024 & 39 & 1 & 2 & 3 \\
Stage 4 (Sentences)        & 1024 & 51 & 1 & 2 & 3 \\
Stage 5 (Extended Context) & 1024 & 63 & -- & -- & 3 \\
\midrule
\textbf{Total stGRU Layers} & \multicolumn{2}{c}{} & \textbf{4} & \textbf{8} & \textbf{15} \\
\bottomrule
\end{tabular}
\end{table}

\subsection{Streaming Implementation via Asymmetric Padding}
To enable the 320\,ms look-ahead streaming variant evaluated in Section \ref{sec:experiments}, we bound future context. Following the $4\times$ spatio-temporal subsampling of the 50\,Hz acoustic features, our pre-encoder operates at an internal frame rate of 12.5\,Hz, where each discrete temporal step represents exactly 80\,ms of audio. Standard temporal convolutions utilize symmetric padding, inherently violating streaming constraints by continually pulling from future acoustic frames. To resolve this, we adopt an asymmetric padding strategy. By eliminating right-sided padding across the majority of the network and deliberately restricting it to accumulate exactly 4 future frames at specific intermediate layers, we bound the network's total future receptive field to exactly 320\,ms ($4 \times 80$\,ms). This bounds algorithmic look-ahead compliance without altering the core matrix-MAC-free MLGRU dynamics.

\section{Reproducibility, Training Infrastructure, and Hyperparameters}
\label{app:hyperparameters}

All model training and empirical evaluations were executed on the LUMI supercomputer cluster using a single LUMI-G node equipped with 4 AMD Instinct MI250X modules, which expose 8 Graphics Compute Dies (GCDs) addressed as independent devices by the runtime. The execution pipeline was containerized using Singularity. Distributed Data Parallel (DDP) training was conducted across the 8 GCDs using \texttt{torchrun} with the \texttt{c10d} backend.

\begin{table}[h]
\centering
\caption{\textbf{Key hyperparameters for PRF neuronal dynamics, optimization, tokenization, and CTC decoding}.}
\label{tab:hyperparameters}
\vspace{0.15cm}
\resizebox{\textwidth}{!}{
\begin{tabular}{@{}llc@{}}
\toprule
\textbf{Component} & \textbf{Hyperparameter} & \textbf{Value / Initialization} \\ 
\midrule
\multirow{4}{*}{\textbf{PRF Dynamics}} 
 & Membrane Time Constant ($\tau$) & Target $1.0$ (with $\mathcal{N}(0, 0.05)$ jitter on $\tau^{-1}$) \\
 & Resonant Frequency ($\theta$)   & Linear spread $\in [0, \pi/2]$ + $\mathcal{N}(0, 0.02)$ jitter \\ 
 & Integration Step ($dt$)         & Target $1.0$ + $\mathcal{N}(0, 0.05)$ jitter \\
 & Spike Threshold ($V_{thr}$)     & Target $1.0$ + $\mathcal{N}(0, 0.05)$ jitter \\
\midrule
\multirow{3}{*}{\textbf{Optimization}} 
 & SRCA Peak Learning Rate ($\eta_{\max}$)   & $10^{-3}$ \\
 & SRCA Valley Learning Rate ($\eta_{\min}$) & $10^{-5}$ \\
 & QAT Warmup Steps                         & 10,990 steps (10\% of total) \\
\midrule
\multirow{2}{*}{\textbf{Tokenization}} 
 & LibriSpeech-960 & SentencePiece BPE (10,000 Vocabulary) \\
 & Aishell-1       & BERT-base-Chinese tokenizer (21,128-token vocabulary) \\
\midrule
\multirow{3}{*}{\textbf{CTC Decoding}} 
 & Beam Size                      & 512 \\
 & Language Model Weight ($\alpha$) & 0.8 \\
 & Word Insertion Penalty ($\beta$) & 1.2 \\ 
\bottomrule
\multicolumn{3}{p{\textwidth}}{\footnotesize \textit{Note:} All four PRF parameters ($\tau, \theta, dt, V_{thr}$) are defined as trainable variables jointly optimized during backpropagation.} \\
\end{tabular}
}
\end{table}

\section{Algorithmic Derivations and Pseudocode}

\subsection{Quantization-Aware Training (QAT) Warmup Strategy}
\label{app:qat_warmup}
To mitigate the optimization instability and ``quantization shock'' associated with aggressively discretizing a randomly initialized network, we employ a progressive QAT warmup strategy. Let $W$ be the continuous full-precision weight matrix. We compute the scaled weights normalized by the absolute mean $\gamma_W$:
\begin{equation}
    W_{\text{scaled}} = \frac{W}{\gamma_W + \epsilon}, \quad W_{\text{ternary}} = \max\left(-1, \min\left(1, \text{round}(W_{\text{scaled}})\right)\right)
\end{equation}
During forward passes, we construct an effective weight matrix $W_{\text{effective}}$ using a scheduling parameter $\alpha_{\text{qat}} \in [0, 1]$:
\begin{equation}
    W_{\text{effective}} = (1 - \alpha_{\text{qat}}) W_{\text{scaled}} + \alpha_{\text{qat}} W_{\text{ternary}}, \quad \alpha_{\text{qat}} = \min\left(1, \frac{\text{step}}{N_{\text{warmup}}}\right)
\end{equation}
At $\alpha_{\text{qat}} = 0$, the network operates in a purely continuous domain. As $\alpha_{\text{qat}} \to 1$, the network is progressively constrained to ternary representations. Backpropagation passes gradients directly through $W_{\text{effective}}$ via the Straight-Through Estimator (STE) to update the continuous latent weights $W$.

\subsection{Deep Residual Stabilization via Channel-Wise Scaling}
\label{app:rezero}
To ensure stable signal propagation across deep layers without adding matrix multiplications, we introduce a layer-wise learnable parameter vector $\alpha \in \mathbb{R}^C$. Let $\mathcal{F}(X_t)$ represent the discrete transformation of the $l$-th Spiking-Ternary block. The residual update is modulated via element-wise multiplication:
\begin{equation}
    Z_t = X_t + \alpha \odot \mathcal{F}(X_t)
\end{equation}
The vector $\alpha$ is initialized to zero (or $10^{-4}$), preserving identity gradient flow during early epochs. Because $\alpha$ is a 1D vector applied element-wise, it requires only $C$ sparse floating-point multiplications per timestep, completely bypassing dense matrix operations.

\subsection{Semi-Resonant Cosine Annealing (SRCA)}
\label{app:lr_scheduler}
\begin{figure}[h]
    \centering
    \includegraphics[width=0.75\textwidth]{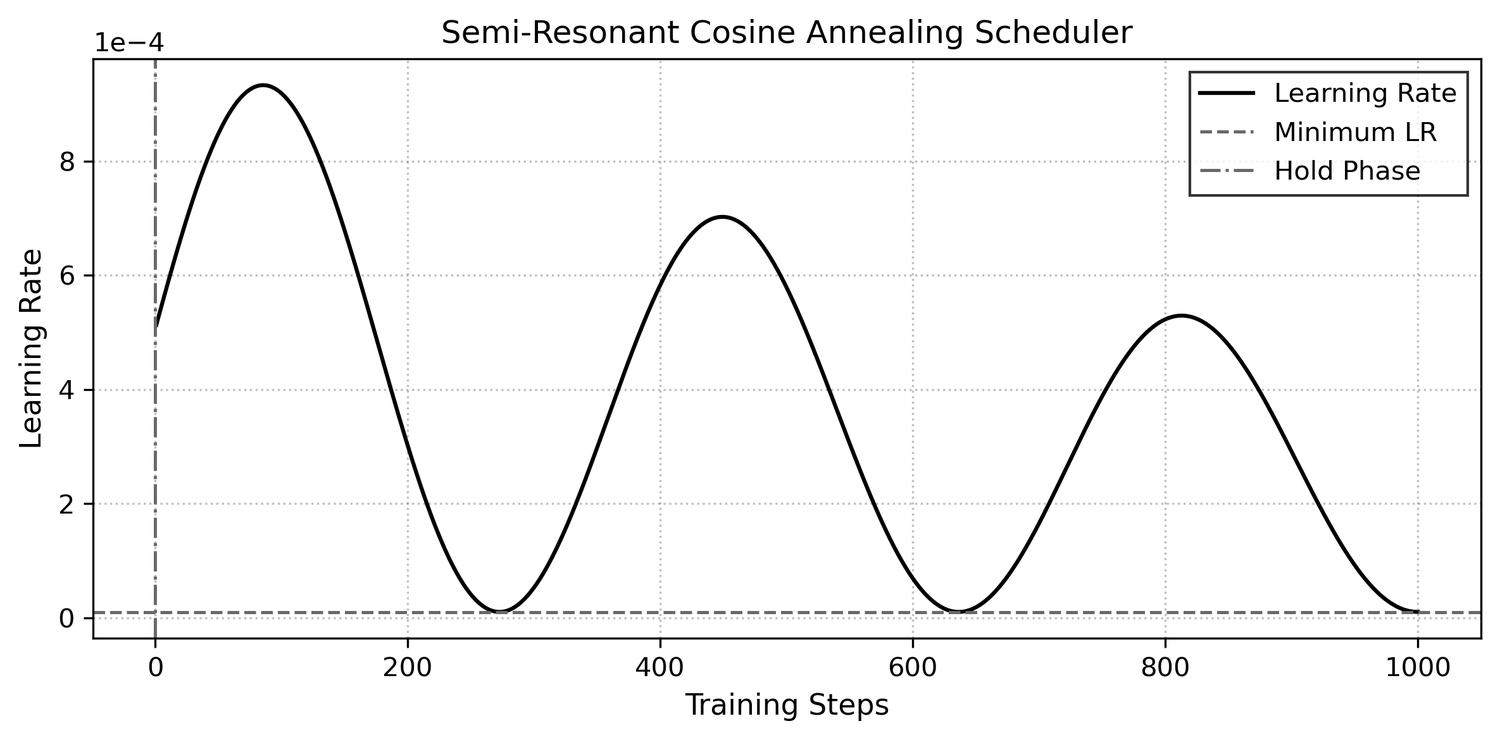}
    \caption{\textbf{Semi-Resonant Cosine Annealing (SRCA) learning rate schedule} with initial holding period ($S_{\text{hold}}=150$), followed by 2.75 cosine cycles with exponential peak decay ($\gamma=0.75$) and initial phase offset ($\phi_{\text{start}}=0.75$).}
    \label{fig:lr_scheduler}
\end{figure}

Let $\eta_{\max}$ and $\eta_{\min}$ denote the peak and anchored valley learning rates. Defining $S_{\text{hold}}$ as the holding period,
$S_{\text{decay}}=S_{\text{total}}-S_{\text{hold}}$, and $c=2.75$ as
the number of cosine cycles, the cycle progress for
$t>S_{\text{hold}}$ is
$C_t=c(t-S_{\text{hold}})/S_{\text{decay}}$. The learning rate $\eta_t$ is governed by:
\begin{equation}
    \eta_t = \begin{cases} 
    \eta_{\max} & \text{if } t \le S_{\text{hold}} \\ 
    \eta_{\min} + \frac{1}{2}(\eta_{\max} - \eta_{\min}) \cdot \gamma^{C_t} \cdot \big(1 + \cos(2\pi (C_t + \phi_{\text{start}}))\big) & \text{if } t > S_{\text{hold}} 
    \end{cases}
\end{equation}
where $\gamma = 0.75$ is the exponential decay factor and $\phi_{\text{start}} = 0.75$ initializes the scheduler at the valley phase for a warmup sweep (Figure \ref{fig:lr_scheduler}).

\begin{algorithm}[H]
\caption{End-to-End Spiking-Ternary Block Forward Pass (with Normalized Residuals)}
\label{alg:deferred_scaling_pipeline}
\begin{algorithmic}[1]
\Require $S^{(l-1)} \in \{0, 1\}^{T \times C}$, Continuous residual $Z^{(l-1)} \in \mathbb{R}^{T \times C}$
\Require Ternary weights $\widetilde{W}$ and structural scales $\gamma$ for all projections
\Require $\alpha, \alpha' \in \mathbb{R}^C$, Initial MLGRU hidden state $H_0$, PRF decay kernel $K_{1:T}$
\vspace{0.1cm}
\State \textit{// 1. Temporal Feature Extraction \& MLGRU Update}
\For{$t = 1$ \textbf{to} $T$}
    \State $X_{dw, t} \leftarrow \mathcal{T}_{dw}(S^{(l-1)}, \widetilde{W}_{dw}, \gamma_{dw})_t$ \Comment{Depthwise matrix-MAC-free operator}
    \State $X_t \leftarrow \mathcal{T}(X_{dw, t}, \widetilde{W}_{pw}, \gamma_{pw})$ \Comment{Pointwise projection}
    \State $C_t \leftarrow \text{SiLU}(\mathcal{T}(X_t, \widetilde{W}_c, \gamma_c))$
\State $F_t \leftarrow \sigma(\mathcal{T}(X_t, \widetilde{W}_f, \gamma_f))$
\State $G_t \leftarrow \text{RMSNorm}(\mathcal{T}(X_t, \widetilde{W}_g, \gamma_g))$
    \State $H_t \leftarrow F_t \odot H_{t-1} + C_t \odot (1 - F_t)$ \Comment{Element-wise parallel scan recurrence}
    \State $O_t \leftarrow \mathcal{T}(G_t \odot \text{SiLU}(H_t), \widetilde{W}_o, \gamma_o)$
\EndFor
\vspace{0.1cm}
\State \textit{// 2. SwiGLU Channel Mixing \& Residual Highway}
\For{$t = 1$ \textbf{to} $T$}
    \State $U'_{t} \leftarrow \text{SiLU}(\mathcal{T}(O_t, \widetilde{W}_{\text{gate}}, \gamma_{\text{gate}})) \odot \mathcal{T}(O_t, \widetilde{W}_{\text{up}}, \gamma_{\text{up}})$
    \State $Z_t \leftarrow \text{RMSNorm}(X_t + \alpha \odot \mathcal{T}(U'_t, \widetilde{W}_{\text{down}}, \gamma_{\text{down}}))$
    \State $Z^{(l)}_t \leftarrow \text{RMSNorm}(Z^{(l-1)}_t) + \alpha' \odot Z_t$
\EndFor
\vspace{0.1cm}
\State \textit{// 3. Parallel Resonate-and-Fire (PRF) Somatic Integration}
\State $U_{1:T} \leftarrow \mathcal{F}^{-1}\big(\mathcal{F}(Z^{(l)}_{1:T}) \odot \mathcal{F}(K_{1:T})\big)$ \Comment{Frequency-domain global parallel integration}
\State $S^{(l)}_{1:T} \leftarrow \mathcal{H}(\Re\{U_{1:T}\} - V_{\text{th}})$ \Comment{Binary spike emission for next block}
\State \Return $S^{(l)}_{1:T}, Z^{(l)}_{1:T}$
\end{algorithmic}
\end{algorithm}

\subsection{Deferred Scaling in Temporal Depthwise Projections}
\label{app:multiplier_free_execution}

For an input $V\in\mathbb R^{T\times C}$, ternary kernel
$\widetilde W_{dw}\in\{-1,0,+1\}^{K\times C}$, and scalar
$\gamma_{dw}$, the naive and deferred forms are
\begin{equation}
\begin{aligned}
X^{\mathrm{naive}}_{t,c}
  &= \sum_k
     (\gamma_{dw}\widetilde W_{dw,k,c})V_{t+k,c},\\
X^{\mathrm{deferred}}_{t,c}
  &= \gamma_{dw}\sum_k
     \widetilde W_{dw,k,c}V_{t+k,c}.
\end{aligned}
\label{eq:hardware_dw_conv}
\end{equation}
They are algebraically equal. In the deferred form, a kernel
coefficient of $+1$, $-1$, or $0$ selects addition, subtraction,
or omission of the corresponding input value. The accumulated
output is then scaled by $\gamma_{dw}$. Input normalization, when
used, precedes both forms and does not distinguish the two modes.
Deferred scaling removes floating-point multiplication by learned
kernel weights inside the temporal sum; it does not remove the
post-accumulation scaling operation.

\section{Hardware-Calibrated Energy and Memory Profiling}
\label{app:profiling_methodology}

\subsection{Dynamic Energy Formulation and Readout Derivation}
\label{app:readout_energy}
Synaptic projection energy scales proportionally with input firing rate $FR \in [0, 1]$ and non-zero weight density $NZ \in [0, 1]$. In standard 45\,nm CMOS technology, floating-point Multiply-Accumulate (MAC) operations consume $E_{\text{MAC}} = 4.6$\,pJ, while floating-point Accumulate (AC) additions consume $E_{\text{AC}} = 0.9$\,pJ \cite{11_horowitz_2014}.

\textbf{MLGRU Gate Energy (over $T$ timesteps):}
\begin{enumerate}
    \item Spike-Ternary Depthwise Conv ($\mathcal{T}_{dw}$): $\mathcal{O}(K \cdot C)$ operation requiring $T \cdot K \cdot C \cdot FR \cdot NZ$ operations at $E_{\text{AC}}$.
    \item Pointwise Conv ($\mathcal{T}$): $\mathcal{O}(C^2)$ operation requiring $T \cdot C^2 \cdot NZ$ operations at $E_{\text{AC}}$.
    \item Gate Projections ($\mathcal{T}$): Candidate and gate representations require $3T \cdot C^2 \cdot NZ$ operations at $E_{\text{AC}}$.
    \item Output Projection ($\mathcal{T}$): Gated recurrent activations require $T \cdot C^2 \cdot NZ$ operations at $E_{\text{AC}}$.
\end{enumerate}

\textbf{SwiGLU Channel Mixing Energy:}
With hidden expansion $E_{\text{swi}} = 4.0$, Up and Gate projections require $8T \cdot C^2 \cdot NZ$ operations, and Down projection requires $4T \cdot C^2 \cdot NZ$ operations, totaling $12T \cdot C^2 \cdot NZ \cdot E_{\text{AC}}$.

\textbf{Neuronal Update Energy:}
Discrete membrane updates per PRF neuron require 4 multiplications and 3 additions per timestep \cite{prf_huang_2024}. Under 45\,nm CMOS ($3.7$\,pJ/mult, $0.9$\,pJ/add), $E_{\text{PRF}} \approx 17.5$\,pJ. Across $C$ channels and $T$ timesteps, this contributes a deterministic $T \cdot C \cdot E_{\text{PRF}}$ cost.

\textbf{Connectionist Temporal Classification (CTC) Readout Energy:}
The CTC linear readout projects the final layer's hidden representations to the vocabulary dimension ($V = 10{,}000$ for LibriSpeech; $V = 21{,}128$ for Aishell-1). Because this projection operates on unquantized full-precision weights $W_{\text{ctc}} \in \mathbb{R}^{V \times C}$ (accounting for the 10.24\,M non-ternary parameter remainder, or 40.96\,MB at FP32), its computational costing requires explicit disambiguation:
\begin{itemize}
    \item \textbf{Input Representation:} The input entering the readout is the binary spike train emitted by the final PRF somatic layer $S^{(L)} \in \{0, 1\}^{T \times C}$.
    \item \textbf{Multiplier-Free Accumulation:} For any output vocabulary token $v \in [1, V]$ at timestep $t$, the logit evaluates as:
    \begin{equation}
        y_{t, v} = \sum_{c=1}^{C} W_{\text{ctc}, v, c} \cdot S^{(L)}_{t, c} = \sum_{c: S^{(L)}_{t, c} = 1} W_{\text{ctc}, v, c}
    \end{equation}
    Because $S^{(L)}_{t, c} \in \{0, 1\}$, scalar weight multiplication by zero is bypassed, and multiplication by one reduces to an identity fetch. Consequently, the ideal sparse implementation of the linear readout fetches
active FP32 weights and accumulates them without floating-point
multiplication. This statement concerns readout arithmetic; weight-fetch
energy is not included in our estimate.
    \item \textbf{Arithmetic Verification:} LibriSpeech \texttt{test-clean} comprises 5.40 hours (19,440\,s). Downsampled at 12.5\,Hz, the evaluation sequence spans $N = 19{,}440 \times 12.5 = 243{,}000$ discrete frames. The dense dimension product is $N \cdot C \cdot V = 243{,}000 \times 1024 \times 10{,}000 \approx 2.488 \times 10^{12}$ theoretical operations. In continuous FP32 baselines, dense activations trigger continuous MACs at $4.6$\,pJ:
    \begin{equation}
        E_{\text{baseline}} = 2.488 \times 10^{12} \times 4.6\,\text{pJ} \approx 11.45\,\text{J} \quad \left(0.59\,\text{mJ/s}\right)
    \end{equation}
    which directly matches the 11.51\,J reported for our parameter-matched FP32 baselines in Table~\ref{tab:energy_profiling_detailed}. In contrast, for our spiking network, the active operation count scales by the empirical final-layer firing rate $FR \approx 42.85\%$ and executes at $E_{\text{AC}} = 0.9$\,pJ:
    \begin{equation}
        E_{\text{stGRU}} = \left(2.488 \times 10^{12} \times 0.4285\right) \times 0.9\,\text{pJ} \approx 1.066 \times 10^{12} \times 0.9\,\text{pJ} \approx \mathbf{0.96\,\text{J}} \quad \left(\mathbf{0.05\,\text{mJ/s}}\right)
    \end{equation}
    This confirms the exact derivation of the 0.96\,J figure reported in Tables~\ref{tab:energy_profiling_full} and \ref{tab:energy_profiling_detailed}.
\end{itemize}

\subsection{Theoretical Energy Estimation Methodology for Baselines}
\label{app:baseline_energy_methodology}
To benchmark dynamic energy consumption against continuous ANNs and hybrid SNNs, we dynamically calculate Multiply-Accumulate (MAC) and Accumulate (AC) operations per audio sample following canonical analytical profiling conventions \cite{11_horowitz_2014, revisiting_hou_2024, imlspikeformer_song_2025}. All continuous baselines replace autoregressive generation with an FP32 Connectionist Temporal Classification (CTC) linear readout costing $N \cdot d \cdot V$ MACs \cite{connectionist_graves_2006, revisiting_hou_2024}. The baseline architectures are decomposed as follows:
\begin{itemize}
    \item \textbf{Vanilla Transformer \cite{vaswani2017attention, revisiting_hou_2024}:} The core encoder operations consist of Multi-Head Attention (MHA) and a single Feed-Forward Network (FFN) per block. The MHA mechanism scales with sequence length $N$ and hidden dimension $d$, requiring $4d^2N + 2dN^2$ MACs ($3d^2N$ for $Q, K, V$ projections, $d^2N$ for output projection, and $2dN^2$ for attention context matrix products). The standard FFN (expansion factor 4) requires $8d^2N$ MACs. Front-end acoustic subsampling uses dense strided 2D convolutions.
    \item \textbf{Conformer \cite{gulati2020conformer, revisiting_hou_2024}:} To capture local acoustic context, the Conformer block introduces a depthwise convolution module adding $K \cdot d \cdot N$ MACs per block for kernel size $K$. Furthermore, it adopts the Macaron layout, sandwiching attention between two half-step FFN modules, doubling FFN compute to $16d^2N$ MACs per block.
    \item \textbf{Transformer++ \cite{revisiting_hou_2024}:} Discards 2D convolutional down-sampling in favor of linear frame-stacking ($r \cdot d_{\text{input}} \times d$) and removes internal encoder convolutions, while retaining the Macaron dual-FFN structure ($16d^2N$ MACs).
    \item \textbf{Hybrid Spiking Architectures (IML-Spikeformer) \cite{imlspikeformer_song_2025}:} Features are downsampled via a continuous FP32 2D frontend incurring 53.13\,J (2.73\,mJ/s) of dense MACs. Its spiking encoder uses HD-RepSSA ($ND^2$ ops), ChannelMLP ($8ND^2$), and attention matrix ops ($2N^2D$) scaled across simulation window $T=6$ and average firing rate $R \approx 18\%$ (10.58\,J / 0.54\,mJ/s). However, total inference remains heavily dominated by its dense 6-layer FP32 autoregressive Transformer decoder (32.72\,J / 1.68\,mJ/s). Whereas IML repeats dynamics over $T=6$ simulation timesteps per frame, stGRU operates at $T=1$ without multi-step latency.
\end{itemize}

\begin{algorithm}[H]
\caption{Profiled Arithmetic-Energy Estimate}
\label{alg:energy_profiling}
\begin{algorithmic}[1]
\Require Model $\mathcal{M}$, Dataset $\mathcal{D}$, Process constants $E_{\text{AC}} = 0.9$\,pJ, $E_{\text{PRF}} = 17.5$\,pJ
\Ensure Total Dynamic Energy $E_{\text{total}}$
\State $E_{\text{total}} \leftarrow 0$
\For{each batch $X \in \mathcal{D}$}
    \State Record forward-pass intermediate activations across all layers $\mathcal{L} \in \mathcal{M}$
    \For{each layer $\mathcal{L} \in \mathcal{M}$}
        \If{$\mathcal{L}$ is a Synaptic Ternary Layer}
            \State $NZ \leftarrow \text{count}(\mathcal{L}.weight \neq 0) / \text{total}(\mathcal{L}.weight)$
            \State $FR \leftarrow (\text{input} \in \{0, 1\}) \; ? \; [\text{count}(\mathcal{L}.input = 1) / \text{total}(\mathcal{L}.input)] : 1.0$
            \State $\text{ActiveOps} \leftarrow \text{TheoreticalOps}(\mathcal{L}) \cdot FR \cdot NZ$
            \State $E_{\text{total}} \leftarrow E_{\text{total}} + (\text{ActiveOps} \cdot E_{\text{AC}})$
        \ElsIf{$\mathcal{L}$ is a Neuronal PRF Layer}
            \State $E_{\text{total}} \leftarrow E_{\text{total}} + (\text{total}(\mathcal{L}.output) \cdot E_{\text{PRF}})$
        \EndIf
    \EndFor
\EndFor
\State \Return $E_{\text{total}}$
\end{algorithmic}
\end{algorithm}

\begin{table}[h]
\centering
\caption{\textbf{Estimated arithmetic-energy decomposition on LibriSpeech test-clean
(5.40 hours; 19,440\,s).} For the continuous ANN comparisons, the downstream column is a modeled
FP32 CTC readout; for IML-Spikeformer it is a six-layer ANN decoder;
for stGRU it is active-weight accumulation in an FP32 CTC readout
driven by binary spikes; see Appendix~\ref{app:readout_energy}). Estimates omit normalization, activation functions, residual scaling,
memory traffic, and decoding/search.}
\label{tab:energy_profiling_full}
\vspace{0.15cm}
\resizebox{\textwidth}{!}{
\begin{tabular}{lccccccccc}
\toprule
\multirow{2}{*}{\textbf{Architecture}} & \multicolumn{4}{c}{\textbf{Estimated Arithmetic Energy (Joules over 19,440\,s)}} & \multicolumn{4}{c}{\textbf{Estimated Arithmetic-Energy Rate (mJ/s)}} & \multirow{2}{*}{\textbf{Dense Matrix MACs}} \\ \cmidrule(lr){2-5} \cmidrule(lr){6-9}
 & \textbf{Sub-Sampler} & \textbf{Encoder} & \textbf{Decoder} & \textbf{Total} & \textbf{Sub-Sampler} & \textbf{Encoder} & \textbf{Decoder} & \textbf{Total} & \\ 
\midrule
\multicolumn{10}{c}{\textit{Continuous Artificial Neural Networks (ANNs)}} \\
\midrule
Transformer 112M \cite{revisiting_hou_2024}   & 53.13 & 117.01 & 1.16 & 171.31 & 2.73 & 6.01 & 0.06 & 8.80 & Yes \\
Transformer++ 112M \cite{revisiting_hou_2024} & 0.18  & 167.73 & 1.16 & 169.08 & 0.009 & 8.623 & 0.06 & 8.69 & Yes \\
Conformer 136M \cite{revisiting_hou_2024}     & 53.13 & 165.19 & 1.16 & 219.49 & 2.73 & 8.49 & 0.06 & 11.28 & Yes \\
\midrule
\multicolumn{10}{c}{\textit{Hybrid Spiking Neural Networks (SNNs) \& Dense ANN Decoder}} \\
\midrule
IML-Spikeformer \cite{imlspikeformer_song_2025} & 53.13 & 10.58 & 32.72 & 96.45 & 2.73 & 0.54 & 1.68 & 4.96 & Yes (Hybrid) \\
\midrule
\multicolumn{10}{c}{\textit{Ours (Ternary SNNs)}} \\
\midrule
\textbf{Ours Small} (Conv1D $\rightarrow$ MLGRU $\rightarrow$ RF)  & 27.46 & 9.86  & 0.92 & 38.24 & 1.41  & 0.50  & 0.04 & \textbf{1.95} & \textbf{No} \\
\textbf{Ours Medium} (Conv1D $\rightarrow$ MLGRU $\rightarrow$ RF) & 27.03 & 20.58 & 0.94 & 48.55 & 1.39  & 1.05  & 0.05 & 2.49          & \textbf{No} \\
\textbf{Ours Large} (Conv1D $\rightarrow$ MLGRU $\rightarrow$ RF)  & 26.73 & 39.01 & 0.96 & 66.70 & 1.374 & 2.006 & 0.05 & \textbf{3.42} & \textbf{No} \\
\bottomrule
\end{tabular}
}
\end{table}

\begin{table*}[h]
\centering
\caption{\textbf{Stage-by-Stage Dynamic Energy Profiling across Model Scales and Architectures for LibriSpeech and Aishell-1}.}
\label{tab:energy_profiling_detailed}
\vspace{0.15cm}
\resizebox{\textwidth}{!}{
\begin{tabular}{@{}lllccccccccc@{}}
\toprule
\multirow{2}{*}{\textbf{Scale}} & \multirow{2}{*}{\textbf{Architecture Variant}} & \multirow{2}{*}{\textbf{Prec.}} & \multirow{2}{*}{\begin{tabular}[c]{@{}c@{}}\textbf{Memory} \\ \textbf{(MB)}\end{tabular}} & \multicolumn{4}{c}{\textbf{Estimated Arithmetic Energy (Joules)}} & \multicolumn{4}{c}{\textbf{Estimated Arithmetic-Energy Rate (mJ/s)}} \\ \cmidrule(lr){5-8} \cmidrule(l){9-12}
 &  &  &  & \textbf{Sub} & \textbf{Enc} & \textbf{Dec} & \textbf{Total} & \textbf{Sub} & \textbf{Enc} & \textbf{Dec} & \textbf{Total} \\
\midrule
\multicolumn{12}{c}{\textit{Dataset: LibriSpeech test-clean (5.40 Hours | 19,440\,s)}} \\
\midrule
\multirow{6}{*}{\textbf{Small}} 
 & Linear $\rightarrow$ RF & FP32$^\ddagger$ & 70.79 & 24.62 & 0.22 & 0.82 & 25.66 & 1.26 & 0.01 & 0.04 & 1.31 \\
 & Conv1D $\rightarrow$ RF & FP32$^\ddagger$ & 71.28 & 26.25 & 4.72 & 0.86 & 31.83 & 1.35 & 0.24 & 0.04 & 1.63 \\
 & MLGRU $\rightarrow$ ReLU & FP32 & 513.32 & 338.47 & 75.48 & 11.51 & 425.46 & 17.40 & 3.88 & 0.59 & 21.87 \\
 & MLGRU $\rightarrow$ RF & INT2 & 70.98 & 24.02 & 9.20 & 0.87 & 34.09 & 1.23 & 0.47 & 0.04 & 1.74 \\
 & Conv1D $\rightarrow$ MLGRU $\rightarrow$ ReLU & FP32 & 530.63 & 338.47 & 80.34 & 11.51 & 430.32 & 17.40 & 4.13 & 0.59 & 22.12 \\
 & \textbf{Conv1D $\rightarrow$ MLGRU $\rightarrow$ RF} & \textbf{INT2} & 72.09 & 27.46 & 9.86 & 0.92 & 38.24 & 1.41 & 0.50 & 0.04 & \textbf{1.95} \\ 
\midrule
\multirow{6}{*}{\textbf{Medium}} 
 & Linear $\rightarrow$ RF & FP32$^\ddagger$ & 87.69 & 23.22 & 0.36 & 0.84 & 24.42 & 1.19 & 0.02 & 0.04 & 1.25 \\
 & Conv1D $\rightarrow$ RF & FP32$^\ddagger$ & 88.69 & 26.41 & 9.44 & 0.88 & 36.73 & 1.35 & 0.48 & 0.04 & 1.87 \\
 & MLGRU $\rightarrow$ ReLU & FP32 & 782.01 & 338.47 & 150.97 & 11.51 & 500.95 & 17.40 & 7.76 & 0.59 & 25.75 \\
 & MLGRU $\rightarrow$ RF & INT2 & 88.09 & 25.32 & 19.36 & 0.89 & 45.57 & 1.30 & 0.99 & 0.04 & 2.33 \\
 & Conv1D $\rightarrow$ MLGRU $\rightarrow$ ReLU & FP32 & 816.63 & 338.47 & 160.68 & 11.51 & 510.66 & 17.40 & 8.26 & 0.59 & 26.25 \\
 & \textbf{Conv1D $\rightarrow$ MLGRU $\rightarrow$ RF} & \textbf{INT2} & 90.31 & 27.03 & 20.58 & 0.94 & 48.55 & 1.39 & 1.05 & 0.05 & \textbf{2.49} \\ 
\midrule
\multirow{6}{*}{\textbf{Large}} 
 & Linear $\rightarrow$ RF & FP32$^\ddagger$ & 117.28 & 24.00 & 0.57 & 0.88 & 25.45 & 1.23 & 0.03 & 0.04 & 1.30 \\
 & Conv1D $\rightarrow$ RF & FP32$^\ddagger$ & 119.56 & 24.77 & 17.70 & 0.93 & 43.40 & 1.27 & 0.91 & 0.04 & 2.22 \\
 & MLGRU $\rightarrow$ ReLU & FP32 & 1252.23 & 338.47 & 283.07 & 11.51 & 633.05 & 17.40 & 14.55 & 0.59 & 32.54 \\
 & MLGRU $\rightarrow$ RF & INT2 & 118.02 & 27.95 & 36.86 & 0.86 & 65.67 & 1.43 & 1.89 & 0.04 & 3.36 \\
 & Conv1D $\rightarrow$ MLGRU $\rightarrow$ ReLU & FP32 & 1317.55 & 338.47 & 301.40 & 11.51 & 651.38 & 17.40 & 15.49 & 0.59 & 33.48 \\
 & \textbf{Conv1D $\rightarrow$ MLGRU $\rightarrow$ RF} & \textbf{INT2} & 122.22 & 26.73 & 39.01 & 0.96 & 66.70 & 1.374 & 2.006 & 0.05 & \textbf{3.42} \\ 
\midrule\midrule
\multicolumn{12}{c}{\textit{Dataset: Aishell-1 test (10.03 Hours | 36,108\,s)}} \\
\midrule
\multirow{6}{*}{\textbf{Small}} 
 & Linear $\rightarrow$ RF & FP32$^\ddagger$ & 116.41 & 57.55 & 0.40 & 2.90 & 60.85 & 1.59 & 0.011 & 0.08 & 1.68 \\
 & Conv1D $\rightarrow$ RF & FP32$^\ddagger$ & 116.91 & 60.20 & 8.78 & 2.70 & 71.68 & 1.66 & 0.24 & 0.07 & 1.97 \\
 & MLGRU $\rightarrow$ ReLU & FP32 & 558.98 & 629.77 & 140.45 & 45.27 & 815.49 & 17.44 & 3.89 & 1.24 & 22.57 \\
 & MLGRU $\rightarrow$ RF & INT2 & 116.61 & 59.71 & 18.33 & 2.98 & 81.02 & 1.65 & 0.50 & 0.08 & 2.23 \\
 & Conv1D $\rightarrow$ MLGRU $\rightarrow$ ReLU & FP32 & 576.25 & 629.77 & 149.49 & 45.27 & 824.53 & 17.44 & 4.14 & 1.25 & 22.83 \\
 & \textbf{Conv1D $\rightarrow$ MLGRU $\rightarrow$ RF} & \textbf{INT2} & 117.72 & 63.58 & 19.81 & 3.00 & 86.39 & 1.76 & 0.55 & 0.08 & \textbf{2.39} \\ 
\midrule
\multirow{6}{*}{\textbf{Medium}} 
 & Linear $\rightarrow$ RF & FP32$^\ddagger$ & 133.32 & 59.64 & 0.51 & 2.55 & 62.70 & 1.65 & 0.01 & 0.07 & 1.73 \\
 & Conv1D $\rightarrow$ RF & FP32$^\ddagger$ & 134.32 & 60.01 & 17.56 & 2.84 & 80.41 & 1.66 & 0.48 & 0.08 & 2.22 \\
 & MLGRU $\rightarrow$ ReLU & FP32 & 827.64 & 629.77 & 280.90 & 45.27 & 955.94 & 17.44 & 7.77 & 1.25 & 26.46 \\
 & MLGRU $\rightarrow$ RF & INT2 & 133.71 & 59.83 & 36.15 & 3.00 & 98.98 & 1.65 & 1.00 & 0.08 & 2.73 \\
 & Conv1D $\rightarrow$ MLGRU $\rightarrow$ ReLU & FP32 & 862.26 & 629.77 & 298.98 & 45.27 & 974.02 & 17.44 & 8.28 & 1.25 & 26.97 \\
 & \textbf{Conv1D $\rightarrow$ MLGRU $\rightarrow$ RF} & \textbf{INT2} & 135.94 & 62.26 & 39.78 & 3.01 & 105.05 & 1.72 & 1.10 & 0.08 & \textbf{2.90} \\ 
\midrule
\multirow{6}{*}{\textbf{Large}} 
 & Linear $\rightarrow$ RF & FP32$^\ddagger$ & 162.91 & 57.76 & 0.78 & 2.59 & 61.13 & 1.60 & 0.02 & 0.07 & 1.69 \\
 & Conv1D $\rightarrow$ RF & FP32$^\ddagger$ & 165.18 & 59.19 & 32.93 & 2.99 & 95.11 & 1.63 & 0.91 & 0.08 & 2.62 \\
 & MLGRU $\rightarrow$ ReLU & FP32 & 1297.86 & 629.77 & 526.69 & 45.27 & 1201.73 & 17.44 & 14.58 & 1.25 & 33.28 \\
 & MLGRU $\rightarrow$ RF & INT2 & 163.65 & 62.23 & 68.72 & 3.06 & 134.01 & 1.72 & 1.90 & 0.08 & 3.70 \\
 & Conv1D $\rightarrow$ MLGRU $\rightarrow$ ReLU & FP32 & 1363.17 & 629.77 & 560.79 & 45.27 & 1235.83 & 17.44 & 15.53 & 1.25 & 34.23 \\
 & \textbf{Conv1D $\rightarrow$ MLGRU $\rightarrow$ RF} & \textbf{INT2} & 167.84 & 61.22 & 76.29 & 2.95 & 140.46 & 1.69 & 2.11 & 0.08 & \textbf{3.88} \\ 
\bottomrule
\multicolumn{7}{l}{\scriptsize $^\ddagger$ Structural baselines use an
INT2 ternary downsampler and FP32 downstream weights; unmarked FP32
variants also use an FP32 downsampler.} \\
\end{tabular}
}
\end{table*}

\begin{table}[h]
\centering
\caption{\textbf{Comprehensive Architectural Comparison on Aishell-1 Mandarin ASR.} Evaluated on Character Error Rate (CER \%). Memory sizes are theoretical footprints based on parameter counts and precision formats.}
\label{tab:comprehensive_aishell}
\vspace{0.15cm}
\resizebox{\textwidth}{!}{
\begin{tabular}{lrlcccc}
\toprule
\multirow{2}{*}{\textbf{Architecture}} & \multirow{2}{*}{\textbf{Params (M)}} & \multirow{2}{*}{\textbf{Weight Type}} & \textbf{Memory} & \textbf{Energy rate} & \multicolumn{2}{c}{\textbf{CER (\%)}} \\ \cmidrule(lr){6-7} 
 & & & \textbf{(MB)} & \textbf{(mJ/s)} & \textbf{dev} & \textbf{test} \\ 
\midrule
\multicolumn{7}{c}{\textit{Continuous Artificial Neural Networks (ANNs)}} \\
\midrule
VGG-BiLSTM \cite{speechbrain_ravanelli_2022} & 93.30 & FP32 & $373.0^\dagger$ & -- & 9.7 & 10.7 \\
Binary S4D$^*$ \cite{learning_stan_2023}     & 21.00 & FP32 & $84.0^\dagger$  & -- & 10.3 & 11.9 \\
Transformer$^*$ \cite{learning_stan_2023}    & 61.90 & FP32 & $247.6^\dagger$ & -- & 5.1  & 5.6  \\
\midrule
\multicolumn{7}{c}{\textit{Hybrid Spiking Neural Networks (SNNs) \& Dense ANN Decoder/LM}} \\
\midrule
LIF$^*$ \cite{imlspikeformer_song_2025}                                    & 22.96 & FP32 & $91.84^\dagger$  & -- & 13.8 & 15.7 \\
Spiking TCN (+ 6-layer ANN Dec)$^*$ \cite{empirical_bai_2018}              & 23.35 & FP32 & $93.40^\dagger$  & -- & 11.4 & 13.7 \\
GSU (+ 6-layer ANN Dec)$^*$ \cite{towards_hao_2024}                        & 60.74 & FP32 & $242.96^\dagger$ & -- & 10.9 & 12.6 \\
Spike-driven Trans. (+ 6-layer ANN Dec)$^*$ \cite{spikedriven_yao_2023}    & 30.35 & FP32 & $121.40^\dagger$ & -- & 10.2 & 11.9 \\
IML-Spikeformer (+ 6-layer ANN Dec) \cite{imlspikeformer_song_2025}        & 61.89 & FP32 & $247.56^\dagger$ & -- & \textbf{5.2} & \textbf{5.7} \\
\midrule
\multicolumn{7}{c}{\textit{Ours (Structural Baselines, FP32 \& W/O Learned ANN Sequence Decoder)}} \\
\midrule
Large (Linear $\rightarrow$ RF) & 88.42 & FP32$^\ddagger$ & 162.91 & 1.69 & 19.95 & 22.92 \\
Large (Conv1D $\rightarrow$ RF) & 88.98 & FP32$^\ddagger$ & 165.18 & 2.62 & 10.81 & 12.52 \\
\midrule
\multicolumn{7}{c}{\textit{Ours (Parameter-Matched Baselines, Continuous Dense ReLU Activations)}} \\
\midrule
Small (Conv1D $\rightarrow$ MLGRU $\rightarrow$ ReLU)  & 144.06 & FP32 & 576.25  & 22.83 & 10.37 & 11.84 \\
Medium (Conv1D $\rightarrow$ MLGRU $\rightarrow$ ReLU) & 215.56 & FP32 & 862.26  & 26.97 & 11.00 & 12.60 \\
Large (Conv1D $\rightarrow$ MLGRU $\rightarrow$ ReLU)  & 340.79 & FP32 & 1363.17 & 34.23 & 10.62 & 12.24 \\
\addlinespace[2pt]
Small (Conv1D $\rightarrow$ MLGRU $\rightarrow$ ReLU)  & 144.06 & Ternary INT2 & 117.64 & 3.56 & 10.19 & 11.68 \\
Medium (Conv1D $\rightarrow$ MLGRU $\rightarrow$ ReLU) & 215.56 & Ternary INT2 & 135.79 & 4.11 & 10.60 & 12.35 \\
Large (Conv1D $\rightarrow$ MLGRU $\rightarrow$ ReLU)  & 340.79 & Ternary INT2 & 167.58 & 5.18 & 9.98  & 11.42 \\
\midrule
\multicolumn{7}{c}{\textit{Ours (matrix-MAC-free, Ternary Spiking Networks \& W/O Learned ANN Sequence Decoder)}} \\
\midrule
Ours Small (Conv1D $\rightarrow$ MLGRU $\rightarrow$ RF)           & 144.08 & Ternary INT2 & \textbf{117.72} & \textbf{2.39} & 10.49 & 12.36 \\
Ours Medium (Conv1D $\rightarrow$ MLGRU $\rightarrow$ RF)          & 215.60 & Ternary INT2 & 135.94          & 2.90          & 9.66  & 11.44 \\
\textbf{Ours Large (Conv1D $\rightarrow$ MLGRU $\rightarrow$ RF)} & 340.86 & Ternary INT2 & 167.84          & 3.88          & \textbf{9.28} & \textbf{10.88} \\
\bottomrule
\multicolumn{7}{l}{\scriptsize $^\dagger$ Theoretical memory computed from parameter count and format precision.} \\
\multicolumn{7}{l}{\scriptsize $^*$ Reproduced results based on publicly available data. Reported numbers correspond to seed 3407.} \\
\multicolumn{7}{l}{\scriptsize $^\ddagger$ Structural baselines use an
INT2 ternary downsampler and FP32 downstream weights; unmarked FP32
variants also use an FP32 downsampler.} \\
\end{tabular}
}
\end{table}

\begin{table}[h]
\centering
\caption{\textbf{Large Model WER (\%) on LibriSpeech across 6 Random Seeds}.}
\label{tab:large_prf_seed_runs}
\vspace{0.15cm}
\resizebox{\textwidth}{!}{
\begin{tabular}{llccccccc}
\toprule
\textbf{Test Set} & \textbf{4-gram LM} & \textbf{Seed 1} & \textbf{Seed 10} & \textbf{Seed 100} & \textbf{Seed 1000} & \textbf{Seed 10000} & \textbf{Seed 3407} & \textbf{Mean $\pm$ Std} \\
\midrule
\multirow{2}{*}{\textbf{test-clean}} 
 & Without & 6.53 & 6.60 & 6.33 & 6.46 & 6.58 & 6.14 & $\mathbf{6.44 \pm 0.17}$ \\
 & With    & 6.01 & 6.12 & 5.86 & 5.98 & 6.09 & 5.66 & $5.95 \pm 0.16$ \\
\midrule
\multirow{2}{*}{\textbf{test-other}} 
 & Without & 17.49 & 17.74 & 17.18 & 17.74 & 17.55 & 17.28 & $\mathbf{17.49 \pm 0.23}$ \\
 & With    & 16.54 & 16.77 & 16.28 & 16.75 & 16.58 & 16.30 & $16.53 \pm 0.21$ \\
\bottomrule
\end{tabular}
}
\end{table}

\begin{table*}[t]
\centering
\caption{\textbf{Component-Wise Ablation: Model Scale, Architecture, and Weight Precision across LibriSpeech and Aishell-1}. Isolates individual contributions of low-precision weights, sequence gating, and discrete spiking activations. \textbf{Reported numbers correspond to seed 3407.}}
\label{tab:ablation_study}
\vspace{0.15cm}
\resizebox{\textwidth}{!}{
\begin{tabular}{lllccccr}
\toprule
\multirow{2}{*}{\textbf{Scale}} & \multirow{2}{*}{\textbf{Architecture Variant}} & \multirow{2}{*}{\textbf{Precision}} & \multirow{2}{*}{\textbf{Params (M)}} & \multirow{2}{*}{\textbf{Memory (MB)}} & \multicolumn{2}{c}{\textbf{Error Rate (\%)}} & \multirow{2}{*}{\textbf{Energy rate (mJ/s)}} \\ \cmidrule(lr){6-7}
 & & & & & \textbf{test-clean / Dev} & \textbf{test-other / Test} & \\
\midrule
\multicolumn{8}{c}{\textit{Part I: LibriSpeech-960 Evaluation (WER \%)}} \\
\midrule
\multirow{8}{*}{\textbf{Small}} 
 & Linear $\rightarrow$ RF                     & FP32$^\ddagger$    & 65.39  & 70.79  & 19.46 & 35.40 & 1.31 \\
 & Conv1D $\rightarrow$ RF                     & FP32$^\ddagger$    & 65.51  & 71.28  & 14.96 & 30.39 & 1.63 \\
 & MLGRU $\rightarrow$ ReLU                    & FP32    & 128.33 & 513.32 & 14.46 & 28.75 & 21.87 \\
 & MLGRU $\rightarrow$ RF                      & Ternary & 128.35 & 70.98  & 15.95 & 30.87 & 1.74 \\
 & Conv1D $\rightarrow$ MLGRU $\rightarrow$ ReLU & FP32    & 132.66 & 530.63 & 7.25  & 19.14 & 22.12 \\
 & Conv1D $\rightarrow$ MLGRU $\rightarrow$ ReLU & Ternary & 132.66 & 72.01  & 7.74  & 20.26 & 2.60 \\
 & Conv1D $\rightarrow$ MLGRU $\rightarrow$ LIF  & Ternary & 132.67 & 72.07  & 8.18  & 20.65 & 1.87 \\
 & \textbf{Conv1D $\rightarrow$ MLGRU $\rightarrow$ RF} & Ternary & 132.68 & 72.09 & 8.09 & 20.79 & 1.95 \\
\midrule
\multirow{8}{*}{\textbf{Medium}} 
  & Linear $\rightarrow$ RF                     & FP32$^\ddagger$    & 69.61  & 87.69  & 18.38 & 34.72 & 1.25 \\
 & Conv1D $\rightarrow$ RF                     & FP32$^\ddagger$    & 69.86  & 88.69  & 13.20 & 28.10 & 1.87 \\
 & MLGRU $\rightarrow$ ReLU                    & FP32    & 195.50 & 782.01  & 12.87 & 26.87 & 25.75 \\
 & MLGRU $\rightarrow$ RF                      & Ternary & 195.54 & 88.09  & 14.50 & 29.18 & 2.33 \\
 & Conv1D $\rightarrow$ MLGRU $\rightarrow$ ReLU & FP32    & 204.16 & 816.63 & 6.41  & 17.29 & 26.25 \\
 & Conv1D $\rightarrow$ MLGRU $\rightarrow$ ReLU & Ternary & 204.16 & 90.16  & 6.68  & 17.71 & 3.13 \\
 & Conv1D $\rightarrow$ MLGRU $\rightarrow$ LIF  & Ternary & 204.18 & 90.27  & 7.30  & 18.72 & 3.21 \\
 & \textbf{Conv1D $\rightarrow$ MLGRU $\rightarrow$ RF} & Ternary & 204.20 & 90.31 & 7.04 & 18.74 & 2.49 \\
\midrule
\multirow{9}{*}{\textbf{Large}} 
 & Linear $\rightarrow$ RF                     & FP32$^\ddagger$    & 77.01  & 117.28  & 17.34 & 33.64 & 1.30 \\
 & Conv1D $\rightarrow$ RF                     & FP32$^\ddagger$    & 77.58  & 119.56  & 11.64 & 26.20 & 2.22 \\
 & MLGRU $\rightarrow$ ReLU                    & FP32    & 313.06 & 1252.23 & 12.13 & 25.24 & 32.54 \\
 & MLGRU $\rightarrow$ RF                      & Ternary & 313.12 & 118.02  & 13.53 & 28.09 & 3.36 \\
 & Conv1D $\rightarrow$ MLGRU $\rightarrow$ ReLU & FP32    & 329.39 & 1317.55 & 5.83  & 16.11 & 33.48 \\
 & Conv1D $\rightarrow$ MLGRU $\rightarrow$ ReLU & Ternary & 329.39 & 121.96  & 6.49  & 17.40 & 4.10 \\
 & Conv1D $\rightarrow$ MLGRU $\rightarrow$ LIF  & Ternary & 329.43 & 122.15  & 7.05  & 18.36 & 3.21 \\
 & \textbf{Conv1D $\rightarrow$ MLGRU $\rightarrow$ RF} & Ternary & 329.45 & 122.22 & 6.14 & 17.28 & 3.42 \\
 & Large (320ms Look-Ahead)                     & Ternary & 329.45 & 122.22  & 9.11  & 22.13 & 3.51 \\
\midrule\midrule
\multicolumn{8}{c}{\textit{Part II: Aishell-1 Evaluation (CER \%)}} \\
\midrule
\multirow{8}{*}{\textbf{Small}} 
 & Linear $\rightarrow$ RF                     & FP32$^\ddagger$    & 76.79  & 116.41 & 21.20 & 24.51 & 1.68 \\
 & Conv1D $\rightarrow$ RF                     & FP32$^\ddagger$    & 76.92  & 116.91 & 12.68 & 14.77 & 1.97 \\
 & MLGRU $\rightarrow$ ReLU                    & FP32    & 139.74 & 558.98 & 18.32 & 20.67 & 22.57 \\
 & MLGRU $\rightarrow$ RF                      & Ternary & 139.76 & 116.61 & 17.66 & 19.87 & 2.23 \\
 & Conv1D $\rightarrow$ MLGRU $\rightarrow$ ReLU & FP32    & 144.06 & 576.25 & 10.37 & 11.84 & 22.83 \\
 & Conv1D $\rightarrow$ MLGRU $\rightarrow$ ReLU & Ternary & 144.06 & 117.64 & 10.19 & 11.68 & 3.56 \\
 & Conv1D $\rightarrow$ MLGRU $\rightarrow$ LIF  & Ternary & 144.08 & 117.69 & 9.15 & 10.59  & 2.18 \\
 & \textbf{Conv1D $\rightarrow$ MLGRU $\rightarrow$ RF} & Ternary & 144.08 & 117.72 &  10.49 & 12.36  & 2.39 \\
\midrule
\multirow{8}{*}{\textbf{Medium}} 
 & Linear $\rightarrow$ RF                     & FP32$^\ddagger$    & 81.02  & 133.32 & 21.20 & 24.51  & 1.73 \\
 & Conv1D $\rightarrow$ RF                     & FP32$^\ddagger$    & 81.27  & 134.32 & 11.63 & 13.68  & 2.22 \\
 & MLGRU $\rightarrow$ ReLU                    & FP32    & 206.91 & 827.64 & 17.52 & 19.77  & 26.46 \\
 & MLGRU $\rightarrow$ RF                      & Ternary & 206.95 & 133.71 & 18.03 & 20.51  & 2.73 \\
 & Conv1D $\rightarrow$ MLGRU $\rightarrow$ ReLU & FP32    & 215.56 & 862.26 & 11.00 & 12.60  & 26.97 \\
 & Conv1D $\rightarrow$ MLGRU $\rightarrow$ ReLU & Ternary & 215.56 & 135.79 & 10.60 & 12.35  & 4.11 \\
 & Conv1D $\rightarrow$ MLGRU $\rightarrow$ LIF  & Ternary & 215.59 & 135.90 & 9.19 & 10.76   & 2.72 \\
 & \textbf{Conv1D $\rightarrow$ MLGRU $\rightarrow$ RF} & Ternary & 215.60 & 135.94 & 9.66 & 11.44  & 2.90 \\
\midrule
\multirow{8}{*}{\textbf{Large}} 
 & Linear $\rightarrow$ RF                     & FP32$^\ddagger$   & 88.42  & 162.91  & 19.95 & 22.92  & 1.69 \\
 & Conv1D $\rightarrow$ RF                     & FP32$^\ddagger$    & 88.98  & 165.18  & 10.81 & 12.52  & 2.62 \\
 & MLGRU $\rightarrow$ ReLU                    & FP32    & 324.46 & 1297.86 & 16.40 & 18.58  & 33.28 \\
 & MLGRU $\rightarrow$ RF                      & Ternary & 324.53 & 163.65  & 16.69 & 18.98  & 3.70 \\
 & Conv1D $\rightarrow$ MLGRU $\rightarrow$ ReLU & FP32    & 340.79 & 1363.17 & 10.62 & 12.24  & 34.23 \\
 & Conv1D $\rightarrow$ MLGRU $\rightarrow$ ReLU & Ternary & 340.79 & 167.58  & 9.98 & 11.42   & 5.18 \\
 & Conv1D $\rightarrow$ MLGRU $\rightarrow$ LIF  & Ternary & 340.84 & 167.77  & 8.45 & 10.05   & 3.71 \\
 & \textbf{Conv1D $\rightarrow$ MLGRU $\rightarrow$ RF} & Ternary & 340.86 & 167.84 & 9.28 & 10.88  & 3.88 \\
\bottomrule
\multicolumn{7}{l}{\scriptsize $^\ddagger$ Structural baselines use an
INT2 ternary downsampler and FP32 downstream weights; unmarked FP32
variants also use an FP32 downsampler.} \\
\end{tabular}
}
\end{table*}

\begin{table}[h]
\centering
\caption{\textbf{Placement of the weight scale in Small and Large
stGRUs.} All modes share input normalization. Naive folds $\gamma$
into the weights; Proposed applies it after ternary accumulation;
None omits it. Proposed has substantially lower modeled
arithmetic-energy rate than Naive, while omitting $\gamma$ worsens
test-clean WER at both scales. Energy rates use the stated 45\,nm
operation costs and exclude an external language model.}
\label{tab:ablation_deferred_scaling}
\vspace{0.15cm}
\begin{tabular}{llccccc}
\toprule
\multirow{2}{*}{\textbf{Model}} & \multirow{2}{*}{\textbf{Scaling Mode}} & \multicolumn{2}{c}{\textbf{WER (\%)}} & \multicolumn{2}{c}{\textbf{Sparsity (\%)}} & \textbf{Energy rate} \\ \cmidrule(lr){3-4} \cmidrule(lr){5-6}
 & & \textbf{test-clean} & \textbf{test-other} & \textbf{Activation} & \textbf{Weight} & \textbf{(mJ/s)} \\
\midrule
\multirow{3}{*}{\textbf{Small}} 
 & Naive ($X(\widetilde{W}\gamma)^\top$)        & 8.15 & 20.92 & 12.93 & 41.53 & 9.78 \\
 & \textbf{Proposed ($(X\widetilde{W}^\top)\gamma$)} & 8.09 & 20.79 & 12.84 & 41.76 & \textbf{1.95} \\
 & None ($X\widetilde{W}^\top$)                 & 9.44 & 22.61 & 14.44 & 39.81 & 2.21 \\
\midrule
\multirow{3}{*}{\textbf{Large}} 
 & Naive ($X(\widetilde{W}\gamma)^\top$)        & 6.44 & 17.43 & 17.36 & 38.57 & 17.74 \\
 & \textbf{Proposed ($(X\widetilde{W}^\top)\gamma$)} & \textbf{6.14} & \textbf{17.28} & 17.08 & 39.15 & \textbf{3.42} \\
 & None ($X\widetilde{W}^\top$)                 & 7.51 & 19.20 & 19.64 & 38.16 & 3.75 \\
\bottomrule
\end{tabular}
\end{table}
\begin{table}[h]
\centering
\caption{\textbf{Empirical Training Throughput for 10 Epochs on LibriSpeech-960}: PRF vs. LIF on 8 AMD Instinct MI250X GCDs.}
\label{tab:lif_vs_prf_time}
\vspace{0.15cm}
\begin{tabular}{llc}
\toprule
\textbf{Model Size} & \textbf{Neuron Type} & \textbf{Training Time (10 epochs)} \\
\midrule
Small  & LIF & 8h 53m \\
Small  & PRF & \textbf{2h 40m} ($3.3\times$ speedup) \\
\midrule
Medium & LIF & 14h 35m \\
Medium & PRF & \textbf{4h 44m} ($3.1\times$ speedup) \\
\midrule
Large  & LIF & 25h 00m \\
Large  & PRF & \textbf{7h 21m} ($3.4\times$ speedup) \\
\bottomrule
\end{tabular}
\end{table}

\section{Neuromorphic Interpretation and Analysis}
\label{app:lognormal}

\begin{figure*}[h]
    \centering
    \includegraphics[width=\textwidth]{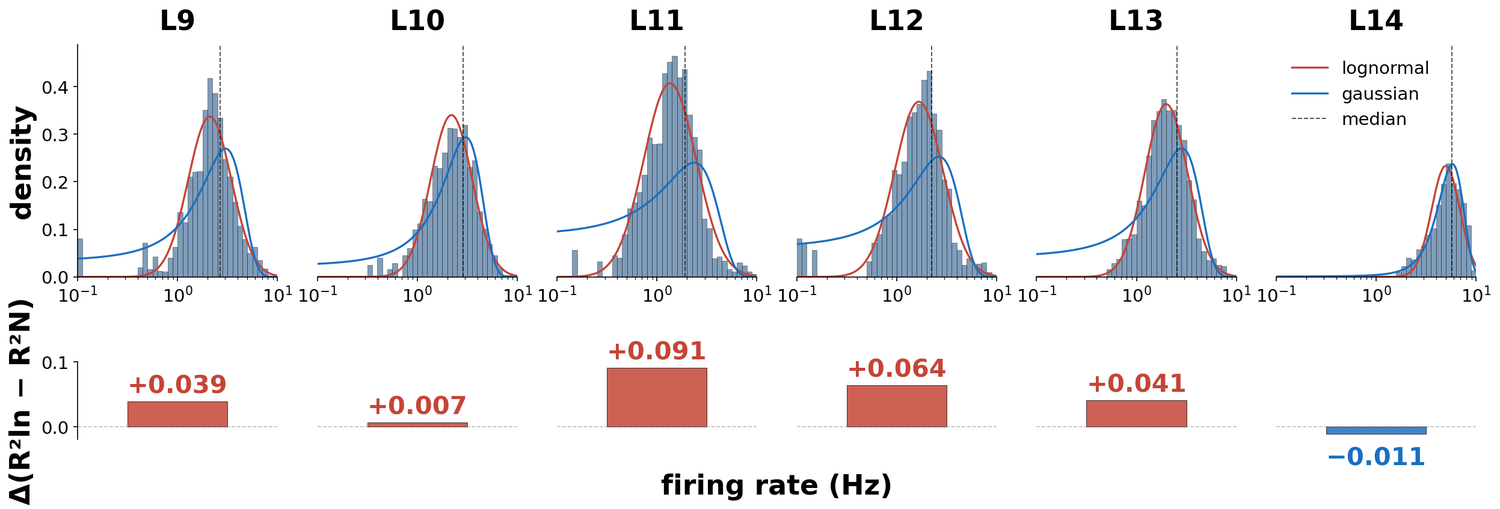}
    \caption{\textbf{Depth-dependent change in firing-rate fits (L9--L14).}
Top: log-axis firing-rate histograms with lognormal (red) and Gaussian
(blue) fits. Bottom: $\Delta R^2=R^2_{\mathrm{ln}}-R^2_{\mathcal N}$;
the lognormal fit has its largest advantage at L11 ($+0.091$), while
the Gaussian fit is slightly better at L14 ($-0.011$).}
    \label{fig:supp_firing_rate_lognormal_vs_gaussian}
\end{figure*}

Per-neuron firing rates were evaluated across the LibriSpeech evaluation set. Neurons firing below 0.1\,Hz were excluded as recording-duration artifacts ($\le 2$ of 1024 neurons per layer; $n=1022$--$1024$ per layer). For each layer, we fit lognormal and Gaussian distributions to empirical firing rates and evaluated $R^2$ as the squared Pearson correlation between empirical and fitted CDFs. We report $\Delta R^2 = R^2_{\text{ln}} - R^2_{\mathcal{N}}$ as the per-layer measure of lognormal superiority. Lognormality peaks at L11 ($+0.091$), persists through L12--L13 ($+0.064$, $+0.041$), and slightly favors Gaussian at L14 ($-0.011$), capturing the transition from an error-monitoring regime to a broad recruitment readout (Figure \ref{fig:supp_firing_rate_lognormal_vs_gaussian}).

\begin{figure*}[h]
    \centering
    \includegraphics[width=\textwidth]{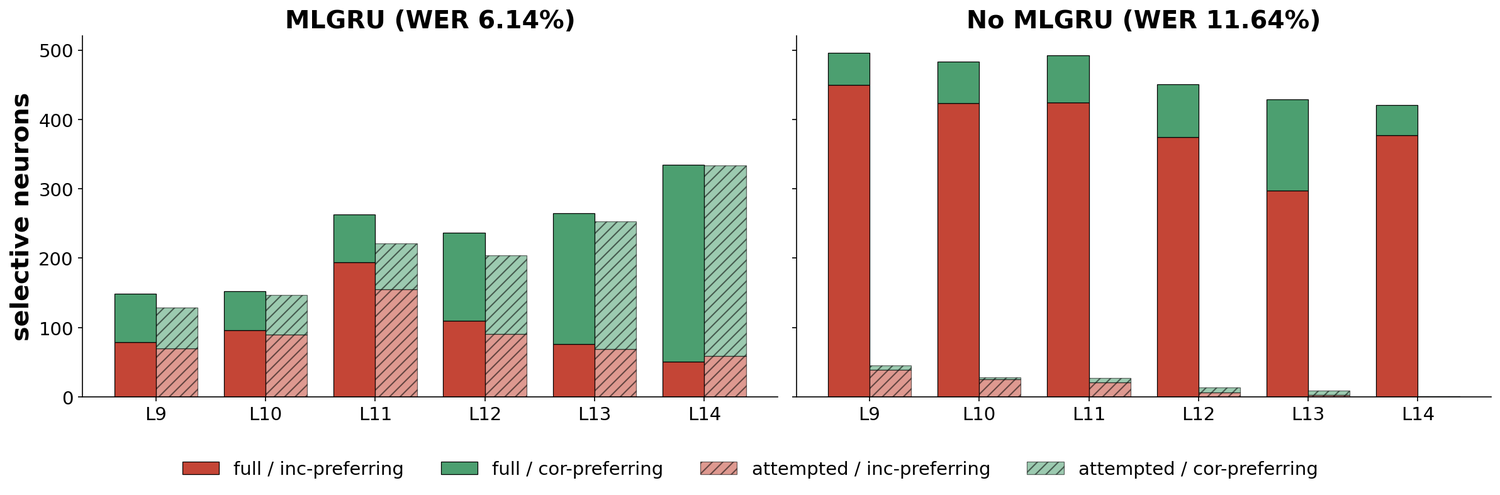}
    \caption{\textbf{Outcome-selective neuron counts (L9--L14) for full vs.\ attempted-emission-only trial pools, MLGRU vs.\ no-MLGRU (Large variants)}. Selective neuron counts per layer, split by outcome preference (incorrect-preferring, red; correct-preferring, green). Left: Full MLGRU (Conv1D $\rightarrow$ MLGRU $\rightarrow$ RF). Empty-emission rate on incorrect words is 13.82\%; restricting to attempted emissions reduces aggregate selective count by only 8.1\%. Right: Ablated variant without MLGRU (Conv1D $\rightarrow$ RF). Empty-emission rate is 57.76\%; restricting to attempted emissions collapses aggregate selective count by 95.5\%, supporting a role for MLGRU gating for the emergent error-monitoring property.}
    \label{fig:mlgru_comparison}
\end{figure*}

\end{document}

%% file: math_commands.tex
\usepackage{amsmath,amsfonts,bm}

\def\eqref#1{equation~\ref{#1}}

\def\1{\bm{1}}

\DeclareMathAlphabet{\mathsfit}{\encodingdefault}{\sfdefault}{m}{sl}
\SetMathAlphabet{\mathsfit}{bold}{\encodingdefault}{\sfdefault}{bx}{n}